%% file: swarmprobe_final.tex
\documentclass{article}
\PassOptionsToPackage{numbers, compress}{natbib}

\usepackage[main,final]{neurips_2026}

\usepackage{graphicx}
\usepackage{subcaption}
\usepackage{enumitem}
\usepackage{amsmath}
\usepackage{multirow}
\usepackage{amssymb}

\usepackage{tabularx}

\usepackage[utf8]{inputenc} 
\usepackage[T1]{fontenc}    
\usepackage{hyperref}       
\usepackage{url}            
\usepackage{booktabs}       
\usepackage{nicefrac}       
\usepackage{microtype}      
\usepackage{xcolor}         

\title{Behavioral Probes for Information Flow in LLM Swarms}

\author{%
Junhui Chen\textsuperscript{1}\thanks{Equal contribution.}\quad
Yu Zhang\textsuperscript{1}\footnotemark[1]\quad
Lantian Li\textsuperscript{1}\thanks{Corresponding author.}\quad
Dong Wang\textsuperscript{2} \\
\textsuperscript{1}School of Artificial Intelligence, \\
Beijing University of Posts and Telecommunications, China \\
\textsuperscript{2}Center for Speech and Language Technologies, BNRist, \\
Tsinghua University, China \\
\texttt{\{rinkovo,rayleighz,lilt\}@bupt.edu.cn} \\
\texttt{wangdong99@mails.tsinghua.edu.cn} \\
\href{https://github.com/HaRry-qaq/Swarmprobe}{\raisebox{-0.25em}[1em][0pt]{\includegraphics[height=1em]{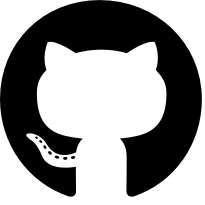}}\ \texttt{github.com/HaRry-qaq/Swarmprobe}}
}

\renewcommand{\acksection}{\section*{Acknowledgment}}

\begin{document}

\maketitle

\begin{abstract}
Most existing approaches to LLM swarm optimization are utility-driven, focusing on final task utility while leaving the processing, integration, and reliance of information within individual interactions largely implicit. To address this, we study LLM swarm interactions from an information-flow-based view and introduce behavioral probes to quantify information flow patterns. Specifically, we repurpose External Context Score (ECS) and Parametric Knowledge Score (PKS)—originally designed for hallucination analysis—to measure a node’s external-context dependence and parametric-knowledge dependence, respectively. Across controlled and realistic swarm settings, we observe consistent patterns: ECS increases with input relevance and distinguishes useful multi-source contributions, while PKS decreases as informative context accumulates. Building on this heuristic interpretation, we demonstrate how these probes can complement utility-driven structure optimization in two controlled applications. Probe-guided node pruning reduces redundant communication and token consumption, and probe-guided reasoning-path search accelerates convergence compared to utility-only baselines. Our empirical results suggest that behavioral probes can make the internal information flow of LLM swarms observable and serve as effective auxiliary signals for structure optimization.

\end{abstract}

\section{Introduction}
Large language model (LLM) swarms have demonstrated strong collaborative reasoning capabilities, with interaction structures evolving from static, hand-designed topologies to learnable directed acyclic graphs (DAGs)~\cite{chen2024survey, chen2025standalone, ferrag2025llm,ke2025survey,zhang2025landscape,guo2024large,yu2025survey,zhu2026survey,wang2025maestro,fan2026todycomm}.

Existing methods for optimizing LLM swarm structures remain largely utility-driven:
edge weights are learned from final task utility~\cite{Yang2025AgentNetDE, zhang2024cut, zhang2026safesieveheuristicsexperienceprogressive},
global rewards are redistributed through Shapley values or heuristic decomposition~\cite{hua2025shapley, li2026deserves, li2026ssvpo, feng2025heterogeneous},
and recent variants incorporate external critiques or explicit topology optimization~\cite{zhang2026reasoningagenticcreditassignment, jiang2025qllm, feng2025group, nagpal2025leveraging, lu2026dytopo, wu2026card, zhang2024g}.
Some methods introduce LLM confidence as intermediate utility signals~\cite{he2025enhancing}
or maintain explicit context states~\cite{chang2025sagallm}.
However, they still do not directly characterize the swarm's internal information flow.
In particular, it remains unclear how information is routed, integrated, and relied upon across interacting nodes.
Thus, individual interactions remain entangled with final task utility,
leaving the swarm's internal information flow largely implicit~\cite{yan2025beyond}.

In this work, we study LLM swarm interactions from an information-flow-based view.
Instead of evaluating swarm structures only through final task utility,
we examine the information-processing behavior behind individual interactions.
To make the information flow formed by these interactions observable,
we instantiate this view with behavioral probes.
Concretely, we use Parametric Knowledge Score (PKS) and External Context Score (ECS),
originally developed for single-model hallucination analysis~\cite{sun2024redeep},
as one concrete but non-unique instantiation of such behavioral probes.
This choice is motivated by their relevance to two behavioral aspects of information processing:
external-context dependence and parametric-knowledge dependence.
Rather than treating this distinction as a formal mechanism,
we use it as a heuristic interpretation for examining information flow across interacting nodes.

Across controlled and realistic swarm settings,
we observe systematic patterns in LLM swarm interactions:
ECS rises with input relevance, segmented ECS distinguishes contributions from multiple upstream sources,
while PKS decreases as informative context accumulates and reasoning converges.
Together, these patterns suggest that behavioral probes can make the information flow formed by individual interactions observable.

These patterns support a heuristic interpretation derived from the observed behaviors:
ECS approximates external-context dependence,
while PKS approximates parametric-knowledge dependence.
This interpretation further suggests two ways to combine behavioral probes with utility signals for structure optimization:
ECS--PKS behavioral profiles can help identify and prune nodes with abnormal behavior,
while PKS-based convergence signals can guide reasoning-path search toward convergent reasoning paths.
It is worth noting that this interpretation is used to examine information flow across interacting nodes,
rather than to provide a formal decomposition of generation.

Building on this heuristic interpretation,
we benchmark behavioral probes in two controlled structure-optimization settings.
In node pruning,
ECS--PKS behavioral profiles reduce redundant communication and token consumption
relative to utility-only baselines.
In reasoning-path search,
probe-based signals consistently identify convergent reasoning paths across three initialization conditions.
In both settings, behavioral probes are combined with utility signals
rather than used as standalone objectives.
In the topology-optimization experiments, metric collection is only needed during diagnosis or training,
while deployment remains standard LLM inference.
Together, these results support the following contributions:

\begin{itemize}[left=0pt, labelsep=5pt, topsep=0pt, partopsep=0pt, itemsep=2pt]

    \item We introduce an information-flow-based view for studying LLM swarm interactions,
    and repurpose ECS and PKS as behavioral probes under this view.
    Across controlled and realistic swarm settings,
    we show that they remain informative for making the information flow formed by individual interactions observable.

    \item We develop a heuristic interpretation of these probes for swarm structure optimization:
    ECS approximates external-context dependence,
    while PKS approximates parametric-knowledge dependence.
    Based on this interpretation,
    we derive two probe-guided ways to complement utility-driven optimization:
    using ECS--PKS behavioral profiles to identify nodes with abnormal behavior,
    and using PKS-based convergence signals to guide reasoning-path search.

    \item We benchmark the probe-guided utility strategies in two controlled structure-optimization settings.
    Together, these results suggest that behavioral probes can guide utility signals in structure optimization:
    ECS--PKS behavioral profiles improve redundant-node pruning with substantially lower token consumption than utility-only baselines,
    while PKS-based signals identify convergent reasoning paths and accelerate the growth of their path weights.
\end{itemize}

\section{Related Work}

\textbf{LLM multi-agent systems.}
LLM-based multi-agent systems have evolved from static, hand-designed topologies~\cite{hong2023metagpt, wu2024autogen, li2023camel, zhang2024can, liu2024dynamic, tian2025agentinit, tekin2025dynamic} toward learnable DAGs~\cite{Yang2025AgentNetDE, feng2025heterogeneous,zhuge2024gptswarm,yang2025topological,qian2024scaling,shen2025understanding}. Yet their optimization remains largely outcome-driven, leaving internal information-flow dynamics undercharacterized.

\textbf{Interaction-level optimization in LLM swarms.}
A key step in optimizing LLM swarm structures is to assign learning signals to individual interactions.
Existing methods remain largely utility-driven:
edge, role, or structural weights are learned from final task utility~\cite{Yang2025AgentNetDE,feng2025heterogeneous},
global rewards are redistributed through Shapley values or heuristic rules~\cite{hua2025shapley,li2026deserves,li2026ssvpo,motwani2024malt},
and recent variants incorporate external critiques or explicit topology optimization~\cite{zhang2026reasoningagenticcreditassignment,jiang2025qllm,feng2025group,nagpal2025leveraging,lu2026dytopo,wu2026card,zhang2024g}.
Process-reward decomposition and context-state modeling introduce intermediate supervision signals~\cite{he2025enhancing,chang2025sagallm},
but they still do not directly characterize the swarm's internal information flow.
In particular, they leave open how information is processed within individual interactions.
In contrast, we use ECS and PKS as behavioral probes under an information-flow-based view,
making the information flow formed by individual interactions observable.

\textbf{Metrics for LLM behaviors.}
Recent work characterizes LLM behaviors related to parametric and contextual knowledge
through discrete knowledge-state modeling~\cite{wang2025continuously, cheng2024understanding},
attribution-based frameworks that distinguish external context from internal parametric knowledge~\cite{sun2024redeep, kahardipraja2025atlas, parry2025mechir,wang2025adacad},
and hallucination detection via internal signals without explicit attribution~\cite{moslonka2026learned, dasgupta2025hallushift, sun2025chair,zhao2025steering,bi2025parameters}.
We adopt PKS and ECS from~\cite{sun2024redeep}
and repurpose them from single-model hallucination analysis to LLM swarm interactions.
This choice is not unique:
other behavioral or attribution metrics could also instantiate an information-flow-based view.
We use ECS and PKS because they are relevant to two behavioral aspects of information processing:
external-context dependence and parametric-knowledge dependence.

\section{Exploratory Analysis of Behavioral Probes in LLM Swarm Structures}
\label{sec:behavioral_signal_analysis}

This section investigates whether ECS and PKS,
repurposed from single-model hallucination analysis~\cite{sun2024redeep},
can remain informative for studying information flow in LLM swarms.
Unlike the original ReDeEP setting, which focuses on whether a single-model retrieval-augmented answer is hallucination-prone,
we ask whether the same signals can reveal how information is processed within individual interactions
and across interacting nodes in LLM swarms.
We conduct experiments across controlled node-level, multi-source aggregation, and chain settings,
as well as a realistic tool-augmented GAIA pipeline.
Together, these settings provide the empirical foundation for interpreting ECS and PKS as behavioral probes.
The illustrated results in the controlled experiments are obtained using
\emph{Qwen2.5-32B-Instruct}~\cite{qwen2};
consistent findings are also observed with
\emph{LLaMA3-70B-Instruct}~\cite{grattafiori2024llama}.

\subsection{Node-Level Response Patterns: Sensitivity to Input Relevance}

Before studying information flow formed by individual interactions in LLM swarms,
we first examine whether ECS and PKS are sensitive to input relevance at a single node.
This calibration establishes a single-node reference point
for later analyses of interactions between nodes.

We design a controlled kinship-question task about public figures, using the open-source data from Berglund et al.~\cite{berglund2023reversal} as the basis for constructing the questions.
Their family relations are unlikely to be directly memorized by the model,
which makes the task suitable for testing sensitivity to input relevance.
For each question, we use four context conditions.
These conditions range from irrelevant context to direct answer evidence: 
\textit{bubble} is unrelated filler context, \textit{rel} is person-related background that does not contain family-relation evidence, \textit{oracle} gives inverse relational evidence sufficient for reverse inference, and \textit{answer} directly states the target relation. This design separates superficial topical relatedness from information that is actually useful for answering. For each question--context pair, we generate one response and compute sequence-level ECS and PKS; detailed dataset construction is provided in Appendix~\ref{app:ch3_node_details}.

\begin{figure}[t]
  \centering
  \begin{subfigure}[b]{0.48\linewidth}
    \centering
    \includegraphics[width=\linewidth]{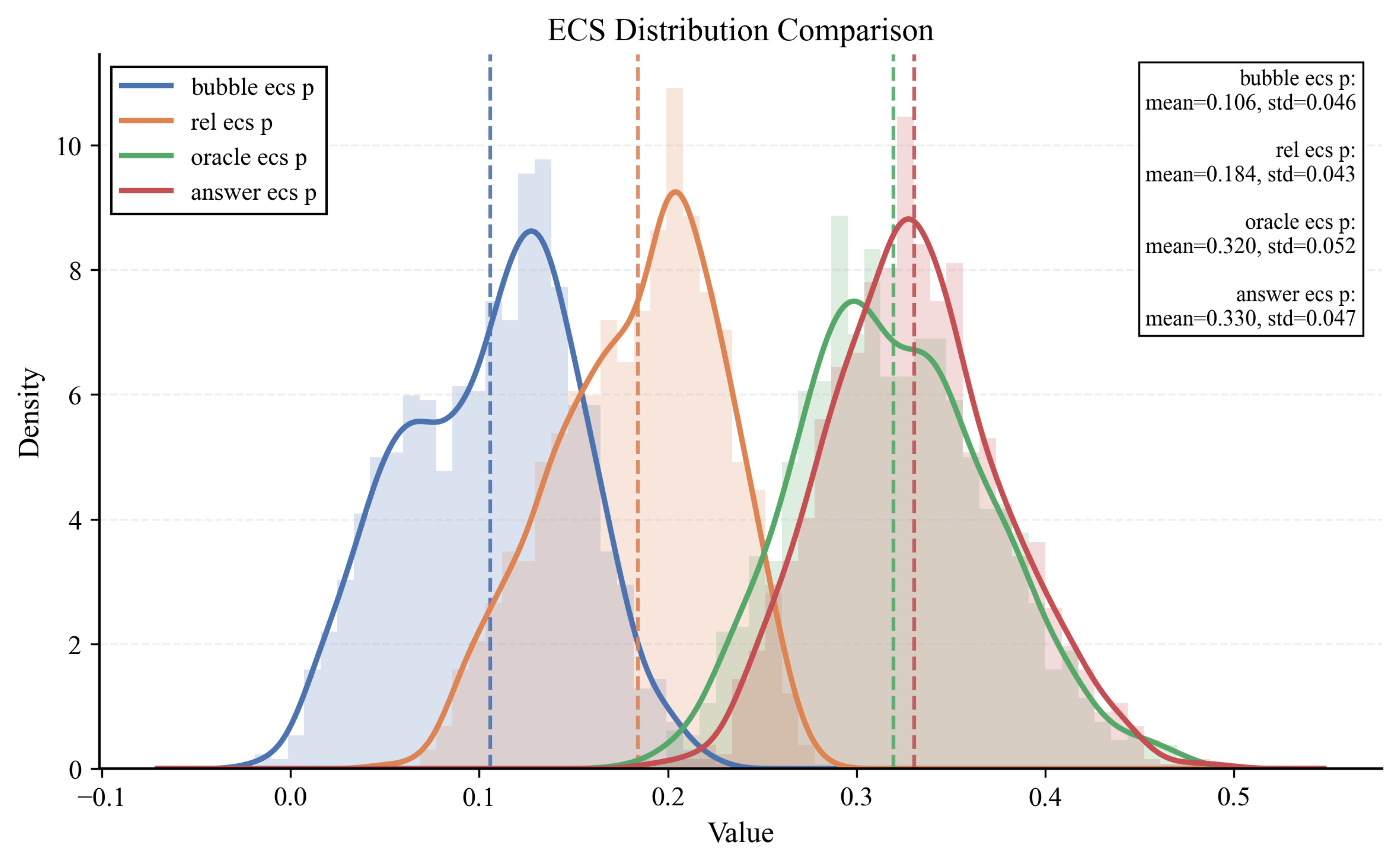}
    \caption{ECS}
    \label{fig:node_ecs}
  \end{subfigure}
  \hfill
  \begin{subfigure}[b]{0.48\linewidth}
    \centering
    \includegraphics[width=\linewidth]{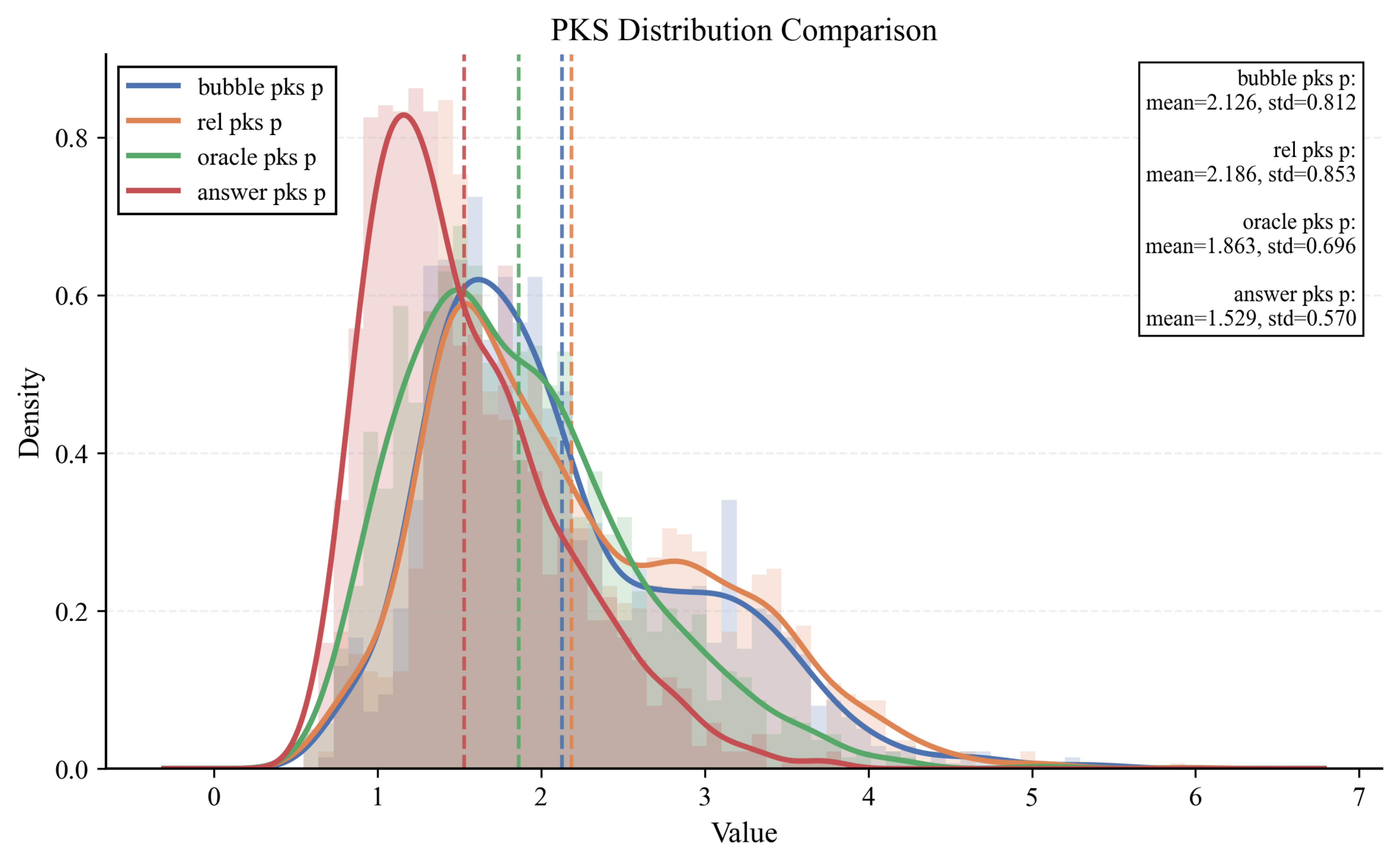}
    \caption{PKS}
    \label{fig:node_pks}
  \end{subfigure}
  \caption{
    Node-level response patterns under four context conditions:
    irrelevant (\textit{bubble}), related but unhelpful (\textit{rel}),
    \textit{oracle} denotes inverse relational evidence that is sufficiently informative for reverse inference, and \textit{answer} denotes context directly containing the answer. Curves and translucent bars show score densities, dashed vertical lines mark group means, and the annotation boxes report mean and standard deviation. Overall, ECS increases with context relevance, while PKS decreases when the answer can be inferred from the provided context.
    }
  \label{fig:node_level}
\end{figure}

As shown in Figure~\ref{fig:node_level},
ECS and PKS show systematic response patterns at the single-node level.
ECS increases with input relevance,
separating answer-supporting conditions (\textit{oracle}, \textit{answer})
from non-answer-supporting ones (\textit{bubble}, \textit{rel}).
It is lowest for unrelated filler context
and highest when the answer is directly provided.
Quantitatively, adjacent relevance levels show consistently higher ECS
(mean adjacent Cliff's $\delta=0.624$, 95\% CI [0.610, 0.640]).
ECS also distinguishes \textit{oracle} from \textit{answer},
suggesting sensitivity beyond a coarse relevant/irrelevant split.

PKS follows the opposite pattern.
It is lowest when the provided context can support the answer,
especially under \textit{answer} and \textit{oracle},
and becomes higher and more dispersed when the context is unrelated
or only provides biographical background.
The adjacent-pair trend is weaker but negative in the expected direction
(mean adjacent Cliff's $\delta=-0.157$, 95\% CI [-0.170, -0.143]).

Together, these results provide a single-node calibration:
ECS is sensitive to useful external context,
while PKS decreases when the answer can be supported by the provided context.
Full effect sizes and statistical tests are provided in Appendix~\ref{app:ch3_node_stats}. Additional decoding and probe-configuration analyses and comparisons with simpler signals are reported in Appendix~\ref{app:additional_checks}.

\subsection{Behavioral Probes in Structured Swarm Interactions}

After the single-node calibration,
we next examine whether ECS and PKS remain informative in structured interactions between nodes.
We focus on two settings that reflect different forms of information flow in LLM swarms:
multi-source aggregation, where a downstream node needs to identify and rely on useful information from multiple upstream sources,
and chain reasoning, where informative context should accumulate across sequential steps.

\subsubsection{Multi-Source Aggregation}

This experiment tests whether ECS can distinguish how a downstream node relies on different upstream sources
when useful evidence is mixed with less useful context.

In LLM swarms, a downstream node often receives concatenated inputs from multiple upstream sources.
The key question is therefore not only whether the overall context is useful,
but also which upstream segments contribute useful information.
Recent work studies related attribution problems with post-hoc methods~\cite{cohen2024contextcite,wang2025tracllm,gao2026atomic}.
Here, we compute segment-level ECS, denoted as \textit{seg-ECS},
as a behavioral probe for examining how the downstream node relies on each upstream segment
in the concatenated context.
It follows the same computation as the original ECS, but restricts the scope from the whole context to a specific context segment.

We instantiate this experiment on a sampled MMLU test set~\cite{hendrycks2020measuring}
using an \textit{Expert $\rightarrow$ Critic $\rightarrow$ FinalDecision} pipeline
with eight expert nodes: seven domain-specific experts and one general Direct Answer expert.
The Critic receives a mixed context composed of their outputs.
These experts are prompt-defined roles using the same model backbone,
rather than agents with separate knowledge bases or tool-calling abilities.
For each question,
one expert domain matches the question category,
while the other experts provide plausible but non-matched content.
This design creates a controlled multi-source setting
where useful and distracting upstream information are presented together,
allowing us to test whether seg-ECS is higher for the relevant upstream source.

\begin{figure}[t]
  \centering
  \begin{subfigure}[b]{0.48\linewidth}
    \centering
    \includegraphics[width=\linewidth]{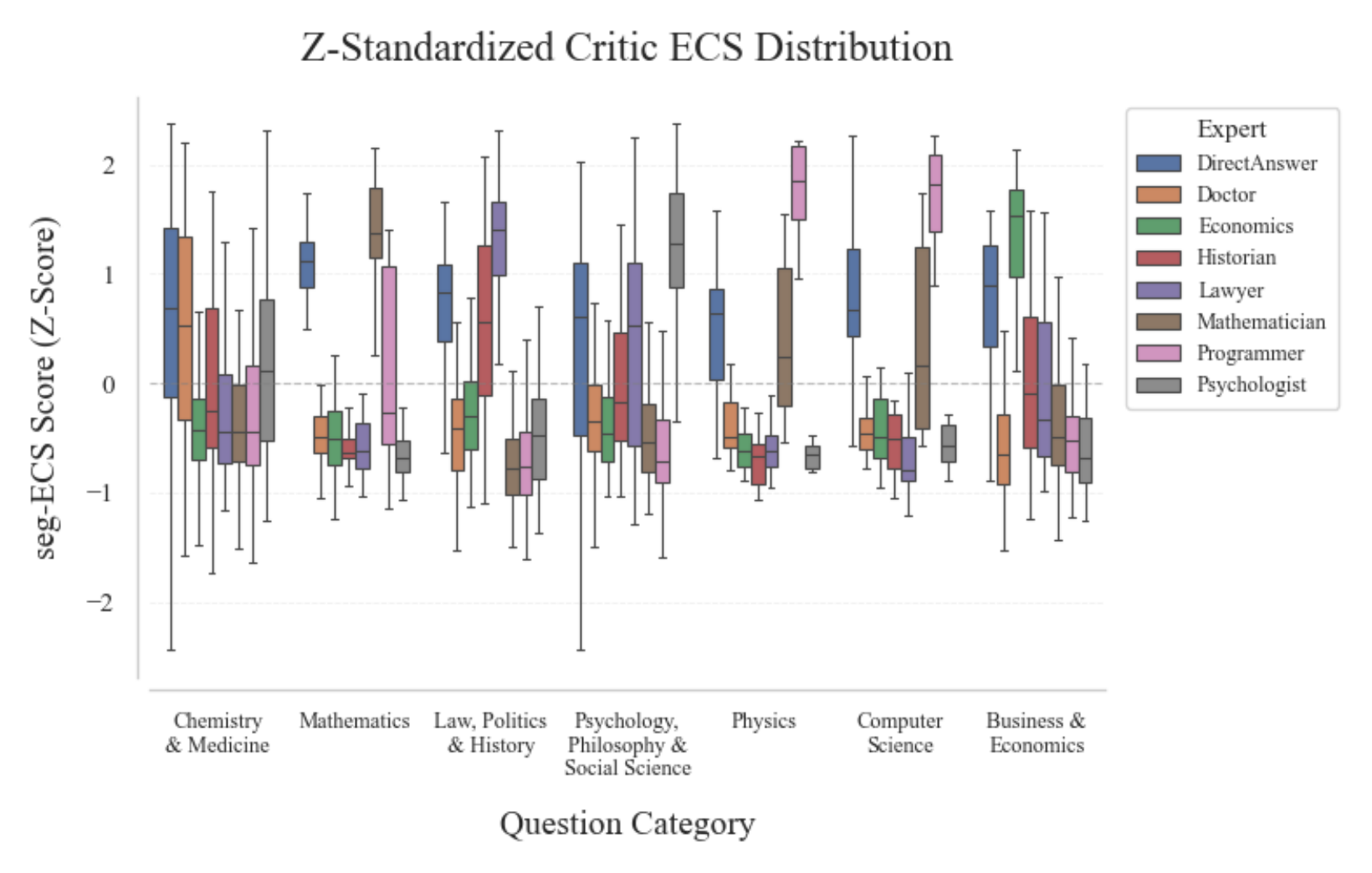}
    \caption{Source-level seg-ECS}
    \label{fig:expert_ecs}
  \end{subfigure}
  \hfill
  \begin{subfigure}[b]{0.48\linewidth}
    \centering
    \includegraphics[width=\linewidth]{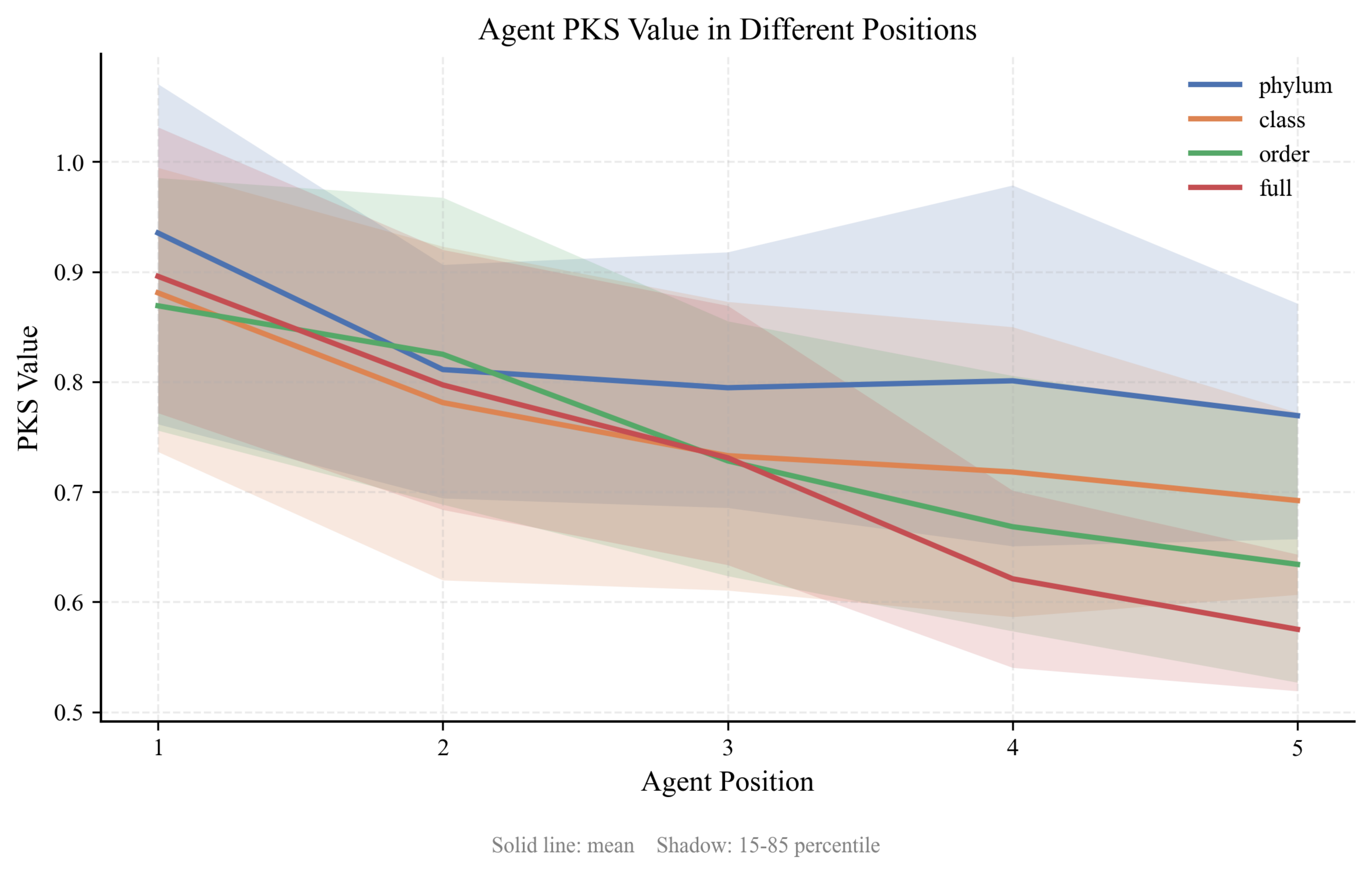}
    \caption{Chain-level PKS trajectory}
    \label{fig:chain_level}
  \end{subfigure}
  \caption{
  Behavioral probes in multi-source aggregation and chain-reasoning settings.
  (a) Z-standardized segment-level ECS of the Critic with respect to upstream expert outputs; higher values indicate stronger reliance on the corresponding segment.
  (b) PKS trajectories along the reasoning chain, with labels indicating the last taxonomic level that introduces new information.
  Solid lines show means and shaded regions show the 15--85 percentile range.
  Overall, seg-ECS highlights domain-relevant upstream sources, while PKS decreases only when informative context continues to accumulate.
  }
  \label{fig:swarm_exp}
\end{figure}

We compute \textit{seg-ECS} for each upstream expert segment.
For each question, we apply z-score normalization across experts,
so that values are comparable across question categories.
Figure~\ref{fig:expert_ecs} also includes the Direct Answer Expert
as a visual reference for a generally useful answer-providing source.
Unlike the domain experts, it attempts to answer each question regardless of the assigned domain role.

As shown in Figure~\ref{fig:expert_ecs},
when an expert domain matches the question category,
the Critic assigns higher \textit{seg-ECS} to that expert than to non-matched experts
(mean paired difference 0.0735, 95\% CI [0.0683, 0.0786]).
This result extends the node-level ECS pattern from input relevance
to source-level contribution in multi-source aggregation.
Detailed experimental setup, filtering criteria, metric computation,
and statistical testing are provided in Appendix~\ref{app:ch3_multi_source}.

\subsubsection{Chain Reasoning and Reasoning Convergence}
\label{sec:chain_reasoning_analysis}

Recent work has explored signals for monitoring reasoning dynamics~\cite{zhao2026entropy,sharma2025think,ni2026reasoning,ye2025uncertainty,zhu2026edis,ghasemabadi2025guided}.
For LLM swarms, this motivates a similar question:
can a behavioral probe track information accumulation along a reasoning path?
We therefore examine whether PKS can reflect reduced parametric-knowledge dependence
as useful context is progressively supplied.

We construct a chain-reasoning setting where the model receives new information at each step.
As the context accumulates,
the model should rely less on parametric knowledge when making predictions.
This experiment tests whether PKS can reflect this change along sequential interactions.

We study PKS dynamics with a biological-taxonomy classification task constructed from NCBI Taxonomy records~\cite{schoch2020ncbi}.
We select 32 species instances and organize their taxonomic information into five rank-like levels:
phylum, class, order, family, and genus.
At each step, the model predicts the genus from the taxonomic information accumulated so far.
To separate effective information gain from chain length,
we use truncated conditions where only the first $k$ steps introduce new taxonomic information ($k=1,2,3,4$),
while later steps repeat the last informative level and continue passing previous outputs along the chain.

As shown in Figure~\ref{fig:chain_level},
PKS decreases monotonically in the full incremental condition,
where each step adds new taxonomic information.
In truncated conditions,
PKS decreases only until the last informative step,
and then stabilizes when later positions repeat the same effective context.
This contrast suggests that PKS responds to information gain,
rather than to chain length alone.
PKS decline should not be interpreted as evidence of correctness:
a lower PKS indicates reduced parametric-knowledge dependence,
but does not guarantee that the provided context is correct
(see Appendix~\ref{app:pks_to_acc} for further discussion).

Together with the multi-source aggregation experiment,
these results show that ECS reflects source-level reliance on useful upstream information,
while PKS reflects information accumulation and reasoning convergence
along structured interactions between nodes.

\subsection{Behavioral Probes in Realistic Tool-Augmented GAIA Reasoning}
\label{sec:gaia_analysis}

We next ask whether these behavioral probes remain informative in a more realistic tool-augmented reasoning pipeline,
where the swarm involves tool calls, intermediate analysis, and final answer synthesis.
This setting extends the analysis beyond controlled structures
while allowing us to examine whether ECS aligns with useful or decision-critical tool evidence
and whether PKS changes when key evidence is withheld from the final answer stage.

We conduct experiments on GAIA Level-1 validation~\cite{mialon2023gaia},
after manually excluding video-modality instances.
Following the original GAIA setup,
we construct a Tree-of-Thought (ToT) reasoning system with four stages.
\textit{GenerateQuery} generates search queries from the input question.
\textit{WebSearch} and \textit{FileAnalyse} are two parallel tool-use branches
that perform function calling and information extraction.
\textit{CombineAnswer} integrates the collected evidence into the final response.

\begin{figure}[t]
  \centering
  \begin{subfigure}[t]{0.32\linewidth}
  \centering
  \includegraphics[width=\linewidth]{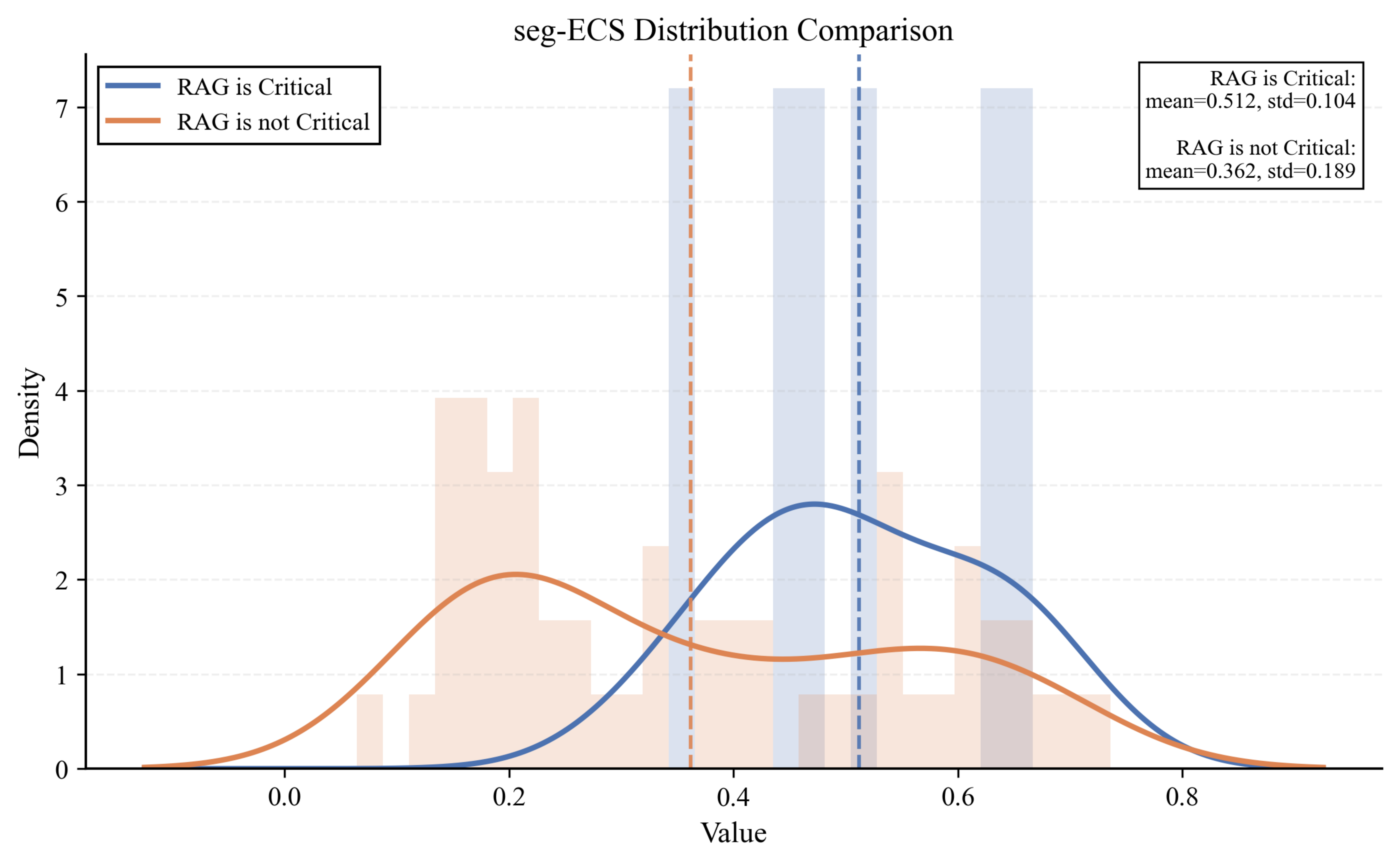}
  \caption{Critical evidence}
  \label{fig:gaia_seg_ecs}
  \end{subfigure}
  \hfill
  \begin{subfigure}[t]{0.32\linewidth}
  \centering
  \includegraphics[width=\linewidth]{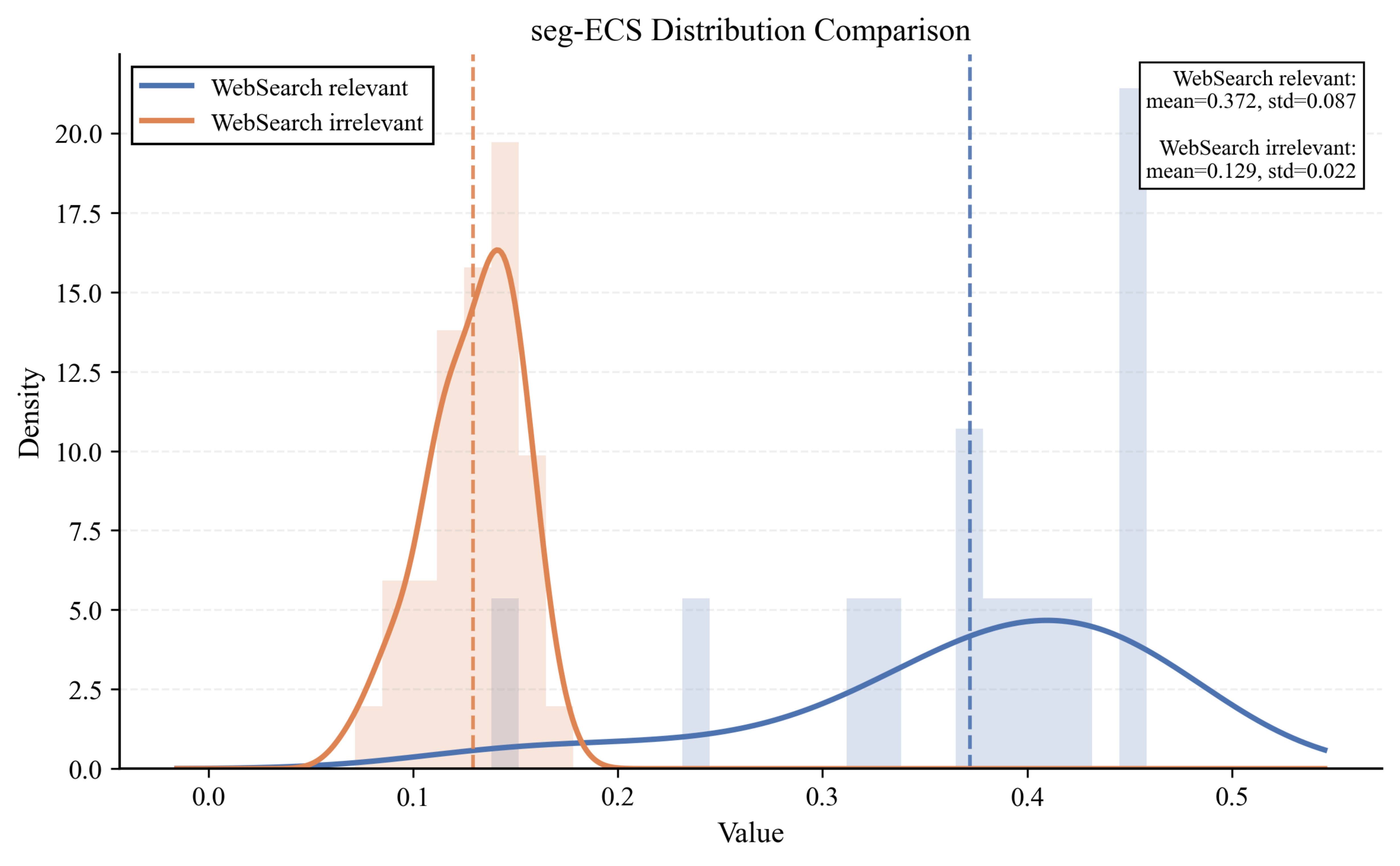}
  \caption{WebSearch relevance}
  \label{fig:gaia_ecs}
  \end{subfigure}
  \hfill
  \begin{subfigure}[t]{0.32\linewidth}
  \centering
  \includegraphics[width=\linewidth]{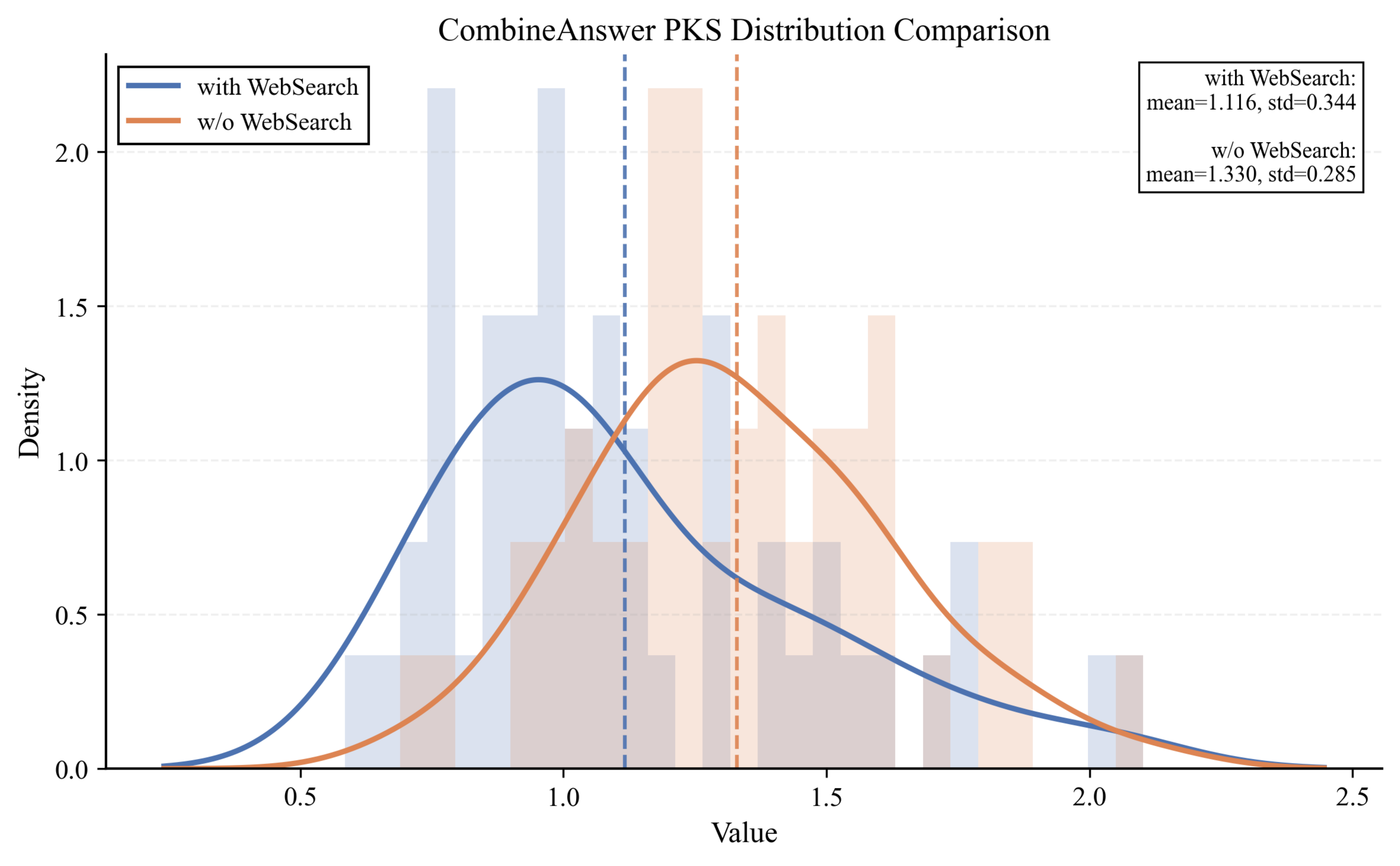}
  \caption{PKS after ablation}
  \label{fig:gaia_pks}
  \end{subfigure}
  \caption{
  Behavioral probes in a realistic tool-augmented GAIA pipeline.
  Panel (a) compares ECS for decision-critical and non-critical tool evidence across \textit{WebSearch} and \textit{FileAnalyse};
  panel (b) compares ECS for function-calling search-result segments labeled as useful or irrelevant by the corresponding \textit{WebSearch} analysis output;
  panel (c) compares \textit{CombineAnswer} PKS before and after removing \textit{WebSearch} evidence.
  }
  \label{fig:gaia_tot}
\end{figure}

We first ask whether ECS aligns with evidence that is decision-critical for final correctness.
In this pipeline, the intermediate outputs from \textit{WebSearch} and \textit{FileAnalyse}
serve as external evidence for \textit{CombineAnswer}.
We define an evidence branch as decision-critical if removing that branch alone changes a correct baseline answer into an incorrect one.
The baseline ToT pipeline achieves 21.2\% accuracy (11/52), and single-branch ablations identify six decision-critical evidence cases:
three from \textit{WebSearch} and three from \textit{FileAnalyse}.
For both branches, removing the entire branch output reduces accuracy from 21.2\% to 15.4\%.
Pruning only high-ECS evidence above the approximately 40th-percentile threshold
($\mathrm{ECS}\approx0.35$) produces the same drop,
whereas pruning low-ECS evidence leaves accuracy unchanged at 21.2\%.
Figure~\ref{fig:gaia_seg_ecs} further shows higher ECS for critical evidence than for non-critical evidence
(mean difference 0.150, 95\% CI [0.052, 0.246]; permutation $p=0.066$).
Given the small number of critical cases,
we treat this result as an exploratory trend rather than a standalone significance claim;
detailed ablation definitions, critical-evidence counts, and statistical tests are provided in Appendix~\ref{app:ch3_gaia_details}.

Beyond decision-criticality,
we test whether ECS aligns with local relevance within \textit{WebSearch}.
Here, segments refer to search results returned by the function-calling tool.
We label each segment using the corresponding \textit{WebSearch} analysis output:
segments whose output contains ``No useful information from WebSearch''
are labeled irrelevant, while the others are treated as useful retrievals.
We then compute segment-level ECS during \textit{WebSearch} generation,
measuring how much the generated analysis relies on each search-result segment.
As shown in Figure~\ref{fig:gaia_ecs},
useful retrieval segments have higher ECS than irrelevant segments,
with a mean difference of 0.243 (95\% CI [0.194, 0.285]).

Finally, we examine whether PKS reflects downstream changes
when external evidence is withheld from the final answer stage.
At the \textit{CombineAnswer} stage,
removing \textit{WebSearch} outputs consistently increases PKS
(Figure~\ref{fig:gaia_pks}),
with a paired mean difference of 0.214 (95\% CI [0.114, 0.309]).
This suggests greater parametric-knowledge dependence when useful external evidence is missing.

Overall, GAIA extends the controlled analysis to a realistic tool-augmented pipeline:
ECS aligns with useful or decision-critical evidence,
while PKS increases when external evidence is withheld from answer synthesis.

\subsection{Summary}
Across controlled and realistic LLM swarm settings,
ECS and PKS exhibit consistent behavioral patterns that make swarm information flow observable.
ECS reflects how downstream nodes use external evidence,
whereas PKS changes as informative context accumulates or is withheld.
These findings motivate the probe-based interpretation developed in the next section.
Detailed experimental configurations and dataset descriptions are provided in Appendix~\ref{app:ch3_details}.

\section{A Heuristic Interpretation of Behavioral Probes for Structure Optimization}
\label{sec:metric_probe_interpretation}
The experiments in Section~3 show that information flow in LLM swarms
---how external context is routed, integrated, and balanced against internal parametric knowledge---
can be characterized through behavioral probes.
ECS rises with input relevance and distinguishes multi-source contributions,
while PKS declines as informative context accumulates and rises when evidence is withheld.
These patterns motivate the heuristic interpretation adopted throughout this work:
\textbf{ECS serves as a heuristic proxy for external-context dependence,
while PKS serves as a heuristic proxy for parametric-knowledge dependence}.
We use this interpretation as an empirically grounded lens for analyzing information flow,
rather than as a formal decomposition of model computation.

Building on this empirical interpretation, we propose that these behavioral probes can complement utility-driven structure optimization in two specific directions.

\textbf{Probe-Guided Node Pruning.} From an information-flow-based view, effective communication requires that nodes both utilize incoming context and contribute non-redundant information to their successors.
Each node can be represented by a behavioral profile $\phi(v) = (ECS_v, PKS_v)$, where ECS is normalized within local aggregation groups and PKS is normalized globally.
Nodes whose profiles deviate from a task-appropriate region $\Omega_{\text{task}}$---for instance,
those that largely ignore incoming messages or act as redundant relays---can be assigned lower structural reward.
Intuitively, $\Omega_{\text{task}}$ represents a ``normal behavior region'' in the ECS--PKS space.
For example, in GAIA, useful \textit{WebSearch} nodes tend to occupy a high-ECS, low-PKS region,
whereas \textit{AdversarialAnswer} nodes fall into a low-ECS, high-PKS region
(visualized in Appendix~\ref{app:ch4_gaia}, Figure~\ref{fig:gaia}).
This profile-based comparison provides a probe-guided view for identifying harmful or superfluous communication paths, complementing utility-driven pruning by characterizing information flow across interacting nodes rather than relying solely on final task utility.

\textbf{Probe-Guided Reasoning-Path Search.} In sequential reasoning, information flow converges when downstream nodes progressively shift from parametric-knowledge dependence toward external-context dependence.
The systematic PKS decline observed in chain reasoning indicates such assimilation of incoming information, providing a behavioral signature of convergence along a communication path.
By tracking these decreases across candidate reasoning edges, one can prioritize paths that move toward convergence within the swarm's sequential information flow.
Since PKS decline reflects perceived information sufficiency rather than factual correctness, we use it as a complementary signal that adds an information-flow dimension to reasoning-path search, not as a substitute for final task utility.

Appendix~\ref{app:ch4_interpretation} provides additional discussion of the assumptions and limitations behind this heuristic interpretation.

\section{Probe-Guided Applications for Structure Optimization}

Building on the heuristic interpretation in Section~\ref{sec:metric_probe_interpretation},
we instantiate two controlled application cases for structure optimization.
Specifically, the node-pruning application uses the ``abnormal ECS--PKS profiles'' rule,
while the reasoning-path search application uses the ``PKS decline as a convergence signal'' rule.

\subsection{Case Study I: Probe-Guided Node Pruning}
\label{sec:node_pruning_case}

The first case uses ECS--PKS behavioral profiles for node pruning.
Beyond final task utility, these profiles characterize two aspects of node behavior:
whether a node's output is used by downstream nodes,
and how much the node itself relies on parametric knowledge.
When ECS is substantially low, the node's information is weakly used by downstream reasoning.
When PKS is substantially low, the node may act as a redundant propagation step after reasoning has already converged.
We therefore use these low-score regions of the ECS--PKS profile
to identify nodes that contribute little useful information flow.

We instantiate this test in a random-DAG pruning task with three node types:
DirectAnswer (DA) nodes answer the question directly,
BubbleAnswer (BA) nodes produce irrelevant answers,
and LazyAnswer (LA) nodes copy existing answers.
Thus, BA outputs should induce low downstream ECS, while LA nodes themselves exhibit low PKS due to their reliance on copied context.
Each swarm contains two DA source nodes, two LA and two BA intermediate nodes,
and one final decision node (FD) as the only sink.
Following the random-DAG construction used in GPTSwarm~\cite{zhuge2024gptswarm},
the optimization space includes all DAG structures satisfying these source, sink, and intermediate-node constraints.
An ideal pruned structure preserves useful DA-to-FD information flow
while suppressing redundant or irrelevant LA/BA contributions.
We evaluate the learned structure by cumulative inbound edge weight to FD,
where higher DA weight and lower LA/BA weight indicate cleaner pruning.

We compare against three utility-driven baselines.
\textit{GPT Swarm}~\cite{zhuge2024gptswarm} is the most basic utility-driven topology optimizer,
which updates the structure using final task utility.
\textit{Agent Prune}~\cite{zhang2024cut} and \textit{Agent Net}~\cite{Yang2025AgentNetDE} inspired baselines also belong to this family,
representing gradient-sensitivity pruning and trajectory-credit assignment, respectively.
In our probe-guided variant,
behavioral probes are combined with utility-based edge updates
as a complementary structural signal.
Detailed experimental settings, including graph construction, evaluation metrics,
and baseline definitions, are provided in Appendices~\ref{app:ch5_details}
and~\ref{app:ch6_baseline}.

\begin{table}[t]
  \centering
  \small
  \setlength{\tabcolsep}{3pt}
  \caption{
  Node-pruning results across baselines.
  DA/LA/BA denote cumulative inbound weights to the final decision node;
  Rdn sums redundant LA/BA weights;
  Tok/Q is average tokens per question.
  The Signal column distinguishes utility-driven baselines from probe-guided utility strategies.
  $^*$ marks variants using ECS--PKS behavioral profiles.
}
  \label{tab:main_results}
  \begin{tabular}{@{} l l l c c c c c c @{}}
  \toprule
  \multirow{2}{*}{Model} & \multirow{2}{*}{Signal} & \multirow{2}{*}{Method}
  & \multicolumn{4}{c}{Weights}
  & \multicolumn{2}{c}{Efficiency} \\
  \cmidrule(lr){4-7} \cmidrule(lr){8-9}
  & & & DA & LA & BA & Rdn & Acc (\%) $\uparrow$ & Tok/Q $\downarrow$ \\
  \midrule
  
  \multirow{6}{*}{Qwen-2.5-32B}
  & \multirow{3}{*}{Vanilla utility}
  & GPT Swarm    & \underline{1.515} & 1.505 & 1.438 & 2.943 & 80.6 & 2294 \\
  & & Agent Net    & \textbf{1.876} & 1.834 & 1.849 & 3.683 & \textbf{80.8} & 2744 \\
  & & Agent Prune  & 1.140 & 0.837 & \underline{0.708} & \underline{1.545} & \textbf{80.8} & 1879 \\
  \cmidrule(lr){2-9}
  & \multirow{3}{*}{Probe-guided utility}
  & Probe-ECS$^*$   & 1.859 & 1.650 & 0.551 & 2.201 & 79.4 & 2345 \\
  & & Probe-PKS$^*$   & 1.052 & \textbf{0.432} & 1.599 & 2.031 & 79.4 & 1710 \\
  & & Probe-Joint$^*$ & 1.474 & \underline{0.491} & \textbf{0.506} & \textbf{0.997} & 80.0 & \textbf{1589} \\
  
  \midrule
  
  \multirow{6}{*}{LLaMA-3-70B}
  & \multirow{3}{*}{Vanilla utility}
  & GPT Swarm    & 1.287 & 1.147 & 1.081 & 2.228 & 72.0  & 2282 \\
  & & Agent Net    & \textbf{1.912} & 1.897 & 1.864 & 3.761 & \textbf{72.2} & 2984 \\
  & & Agent Prune  & 1.262 & 1.111 & 1.068 & 2.179 & 71.4 & 1980 \\
  \cmidrule(lr){2-9}
  & \multirow{3}{*}{Probe-guided utility}
  & Probe-ECS$^*$   & \underline{1.684} & 1.394 & \textbf{0.123} & \underline{1.517} & 71.4 & 2501 \\
  & & Probe-PKS$^*$   & 0.353 & \underline{0.308} & 1.866 & 2.174 & 70.2 & 1691 \\
  & & Probe-Joint$^*$ & 1.528 & \textbf{0.297} & \underline{0.201} & \textbf{0.498} & 71.2 & \textbf{1563} \\
  
  \bottomrule
  \end{tabular}
\end{table}

Table~\ref{tab:main_results} shows that Probe-Joint recovers the cleanest topology across both models.
It achieves the lowest redundant LA/BA weight,
while preserving comparable accuracy and reducing token consumption.
In contrast, utility-driven baselines retain larger redundant LA/BA weights.
This suggests that final task utility alone provides limited guidance
for pruning unnecessary node contributions.
The Agent-Prune-inspired baseline improves sparsity, but the recovered topology is still less clean than that obtained by Probe-Joint.
The single-probe ablations reveal both complementarity and limitation.
Probe-ECS mainly suppresses BA contributions,
consistent with its role in identifying information that is not useful for downstream nodes.
Probe-PKS mainly suppresses LA contributions,
but it can also reduce DA weights.
This occurs because BA is a divergent generation node:
although its output is question-irrelevant,
it can still receive high PKS due to stronger parametric-knowledge dependence.

Probe-Joint addresses this limitation by requiring both behavioral axes to be considered together.
ECS filters question-irrelevant BA contributions,
while PKS identifies redundant LA contributions after the reasoning has already converged.
This highlights the value of behavioral profiles as complementary signals for probe-guided node pruning. A separate paired evaluation on 500 test questions quantifies the accuracy--communication trade-off; few-shot expert selection and conditional escalation experiments are also reported in Appendix~\ref{app:additional_checks}. 

\begin{figure}[t]
  \centering
  \begin{subfigure}[t]{0.32\linewidth}
  \centering
  \includegraphics[width=\linewidth]{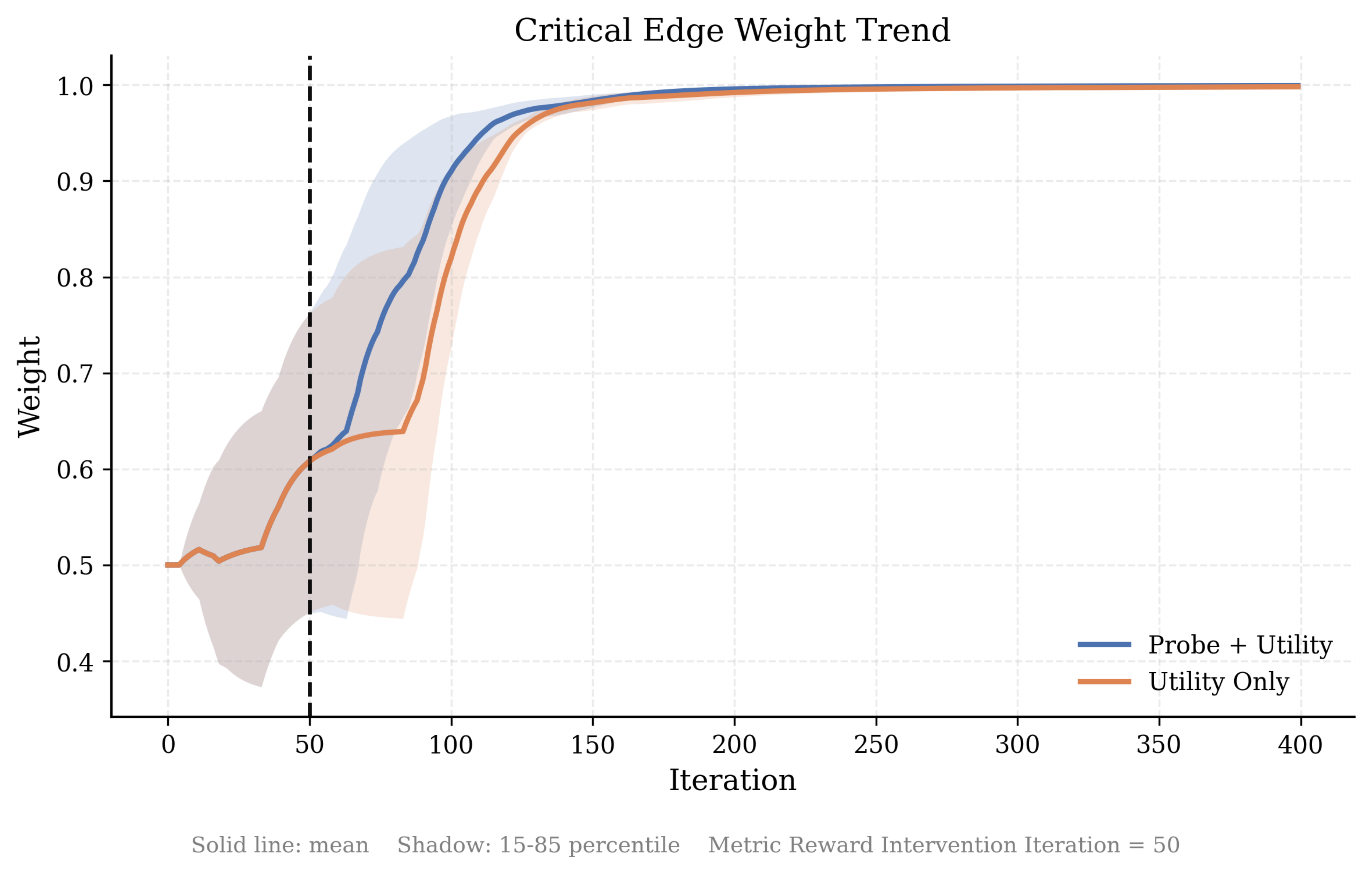}
  \caption{Good warm-starts}
  \label{fig:pks_chain_util_vs_metric}
  \end{subfigure}
  \hfill
  \begin{subfigure}[t]{0.32\linewidth}
  \centering
  \includegraphics[width=\linewidth]{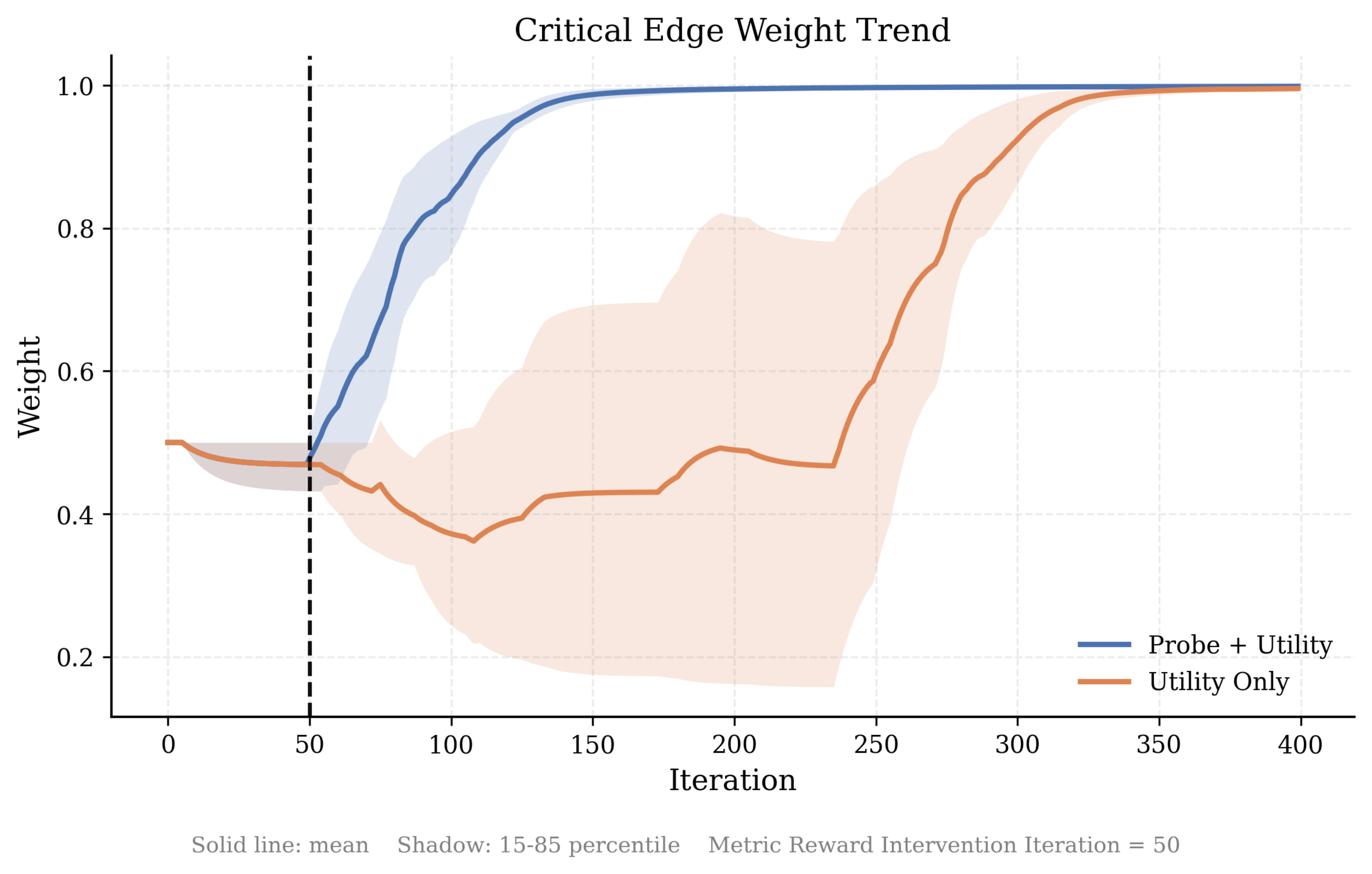}
  \caption{Moderate warm-starts}
  \label{fig:pks_chain_detail}
  \end{subfigure}
  \hfill
  \begin{subfigure}[t]{0.32\linewidth}
  \centering
  \includegraphics[width=\linewidth]{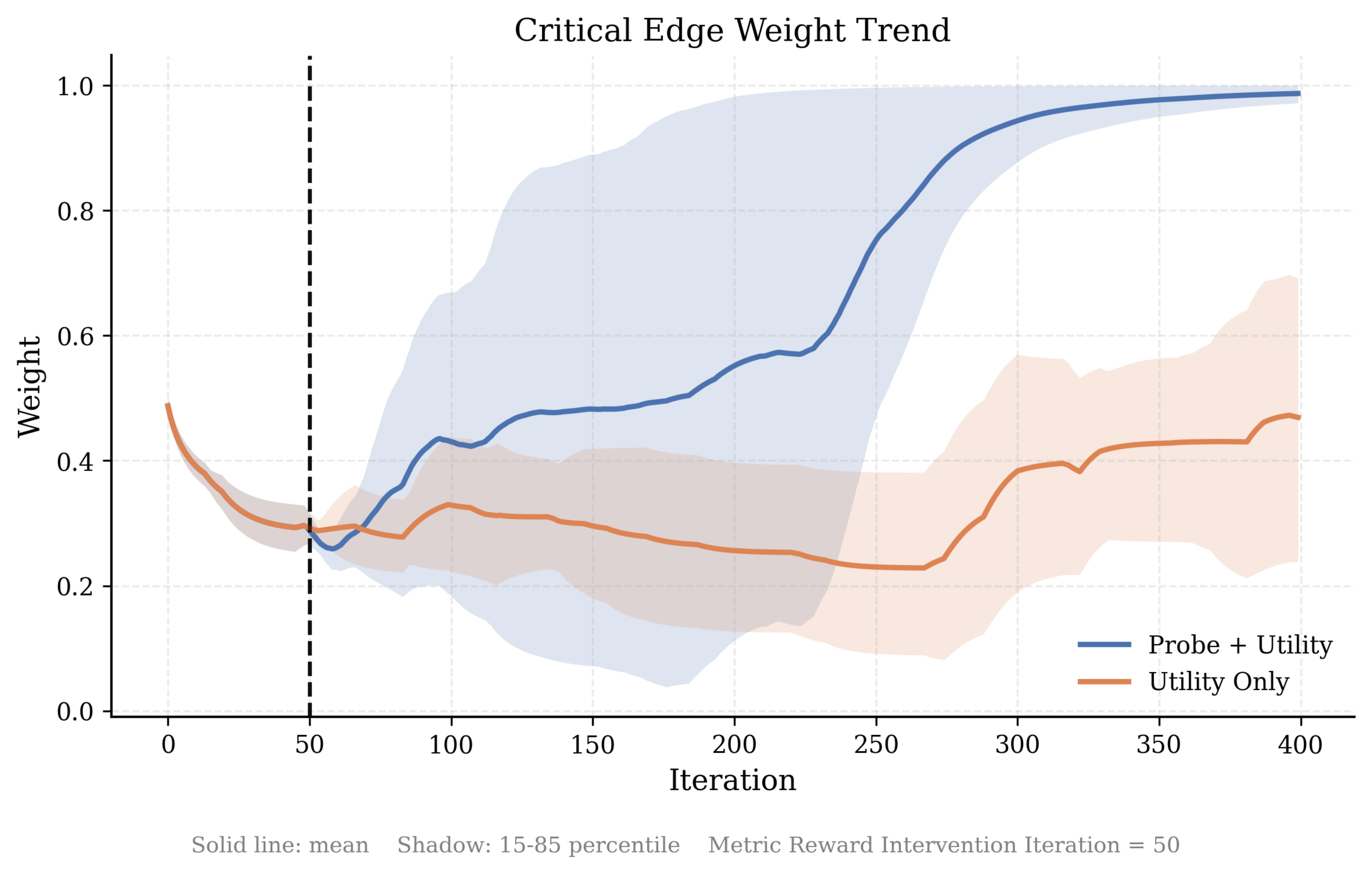}
  \caption{Poor warm-starts}
  \label{fig:pks_chain_only_ml}
  \end{subfigure}
  \caption{
  Critical edge weight trends under three warm-starts.
  After 50 utility-only warm-up iterations, the PKS-based convergence probe is introduced.
  The panels show good, moderate, and poor warm-starts, defined by the critical-edge weight before probe intervention.
  Solid lines denote the mean weight over critical edges, with shaded regions showing the 15--85 percentile range.
  Across all cases, the probe accelerates critical-edge growth compared with the utility-only baseline.
}
  \label{fig:pks_chain_triptych}
\end{figure}

\subsection{Case Study II: Probe-Guided Reasoning-Path Search}
\label{sec:path_selection_case}

The second case uses a PKS-based convergence probe for reasoning-path search.
For each candidate edge,
a larger downstream PKS decrease suggests that the upstream information helps reduce parametric-knowledge dependence.
We therefore use this signal to prioritize convergent paths,
while retaining utility signals to preserve task correctness.

We consider a taxonomic reasoning task similar to the chain analysis in Section~\ref{sec:chain_reasoning_analysis},
but adapt it for reasoning-path search.
The task contains five ordered reasoning nodes corresponding to five taxonomic levels,
and paths are sampled with a softmax-weighted Markov-chain procedure; see
Appendix~\ref{app:taxonomic_labyrinth} for details.

Taxonomic context is accumulated only when the sampled nodes follow the correct order;
otherwise, downstream nodes receive only the original highest-level taxonomic information.
Thus, among all sampled paths, only the fully ordered chain follows the complete taxonomic progression toward the final genus.

This design creates a targeted test for reasoning-path search:
task utility is observed only after a sampled path produces a final genus prediction,
while the ordered taxonomy provides intermediate structure for detecting reasoning convergence.
We therefore examine whether the PKS-based convergence probe can guide path selection toward the correct ordered chain.

Unlike the chain analysis in Section~\ref{sec:chain_reasoning_analysis},
this experiment does not propagate full generated responses between nodes;
downstream nodes receive only structured taxonomic clues,
which keeps intermediate inputs controlled and reduces generation-induced variation.

For reasoning-path search,
we introduce the PKS-based convergence probe after a 50-iteration utility-only warm-up.
During warm-up, PKS is collected on each edge and EMA-smoothed,
allowing edge-level PKS-drop patterns to stabilize before optimization uses them.
We then compare \textit{Probe+Utility}, which adds the PKS-based reward to utility,
against \textit{Utility Only}, which continues using only utility reward.

The probe reward is based on the smoothed PKS drop along each edge,
$\Delta \bar{PKS}_e = \bar{PKS}_u - \bar{PKS}_v$.
A positive drop indicates that the downstream node becomes less dependent on parametric knowledge
after receiving the upstream information.
Edges with larger PKS drops are therefore treated as more likely to lie on convergent reasoning paths.
The detailed reward formulation is provided in Appendix~\ref{app:path_pks_reward}.

Figure~\ref{fig:pks_chain_triptych} compares the two settings under three different warm-starts.
In the good warm-start,
the first 50 iterations have already explored the critical convergent path.
In the moderate warm-start,
the warm-up phase receives little utility reward,
so the weight of the critical edge remains close to its initial level.
In the poor warm-start,
the critical edge receives low weight after warm-up.
Because edge selection is based on softmax sampling,
a wrong path reinforced by utility signals can suppress the sampling probability
of the critical edge,
especially when that wrong path reaches the correct answer by chance.

Across all three cases,
introducing the PKS-based convergence probe after iteration 50 helps identify the critical edge
and accelerates the growth of its weight.
This suggests that the PKS-based convergence probe can provide useful local guidance
when utility signals are sparse, delayed, or misleading.
We observe the same qualitative trends with \emph{LLaMA3-70B-Instruct}~\cite{grattafiori2024llama}.

\section{Conclusion}

This paper studies LLM swarms from an information-flow-based view.
We show that ECS and PKS can serve as behavioral probes for making the information flow formed by individual interactions observable.
Based on this heuristic interpretation,
we further propose two ways to use these probes to complement utility rewards in structure optimization:
ECS--PKS behavioral profiles for node pruning,
and PKS-based convergence signals for reasoning-path search.
The results demonstrate the potential of combining behavioral probes with utility rewards
for LLM swarm structure optimization.

\begin{ack}
This work was supported by the National Natural Science Foundation of China under Grant 62171250 and Grant 62301075.
\end{ack}

\bibliographystyle{unsrtnat}
\bibliography{cite}

\newpage
\appendix

\section{Limitations and Future Work}

Several limitations warrant discussion.
First, computing ECS and PKS requires access to internal model states,
including hidden representations and attention weights.
This incurs additional computational overhead during training
and is not feasible for black-box API-based models.
While this restricts immediate practical deployment,
the information-flow-based view and heuristic interpretation advanced here
remain useful for understanding information flow in LLM swarms
and for motivating future research into more accessible behavioral probes.
Second, our experiments are limited to controlled, small-scale benchmarks;
scalability to complex, open-ended environments remains to be tested.
Third, the reward designs derived from the heuristic interpretation are ad hoc,
and more principled ways of combining behavioral probes with utility signals
may improve performance.
Fourth, ECS and PKS may vary in sensitivity across model architectures and tasks.
Finally, PKS decline reflects perceived information sufficiency rather than correctness,
motivating integration with complementary cues such as uncertainty estimates.
Addressing these limitations is essential for turning the present empirical tests
into practical tools.

These limitations also clarify the intended scope of our claims.
We do not argue that ECS or PKS should replace utility signals,
nor that the observed probe patterns provide causal guarantees about model behavior.
Rather, the results support a narrower claim:
under accessible-model settings where internal states can be measured,
ECS and PKS can serve as behavioral probes
that complement final task utility.
Future work should test whether cheaper proxy signals,
black-box approximations,
or uncertainty-calibrated variants can preserve similar sensitivity to information flow
at larger scale.

Importantly, this paper should be read as an empirical study
rather than a ready-to-deploy method.
Our primary goal is to show that behavioral probes can make meaningful information-flow patterns
observable in LLM swarms.
The two application experiments,
node pruning and reasoning-path search,
serve as controlled tests of downstream possibilities,
not as fully optimized techniques.
Claims of practical superiority over utility-driven optimization
would require extensive evaluation on diverse real-world tasks,
which lies beyond the scope of this initial study.

In addition, ECS and PKS should not be treated as definitive explanations or safety guarantees, especially when such systems are considered for high-stakes use. High ECS can reflect reliance on incorrect evidence; copying or false convergence can also improve probe rewards without improving correctness. Task/role calibration, utility checks, and evidence verification remain necessary. Better diagnosis and efficiency may benefit agent systems, while over-trusting these signals or strengthening harmful automation can have negative impacts.

\section{LLM Swarm Topology and Behavioral Metrics}
\label{app:swarm_definition}

This section formally defines the Directed Acyclic Graph (DAG) representation of LLM swarms,
the optimization objective for structure selection,
and the computational procedures for the External Context Score (ECS) and Parametric Knowledge Score (PKS).
We adapt ECS and PKS from ReDeEP~\cite{sun2024redeep},
but use a modified deterministic layer/head selection strategy informed by its empirical findings on layer and head selection;
the DAG-based swarm optimization setup follows GPTSwarm~\cite{zhuge2024gptswarm}.
\subsection{Swarm Topology as a Directed Acyclic Graph}

In an LLM swarm, the collaborative topology is represented as a DAG. Specifically, an agent or a computational workflow can be formalized as a computational graph:
\begin{equation}
G = (N, E, F, o)
\end{equation}
where $N$ denotes the set of nodes, each corresponding to a computational unit such as an LLM call, a tool invocation, or another processing module; $E \subset N \times N$ denotes the set of directed edges describing the information flow between nodes; $F = \{f_n\}_{n \in N}$ denotes the computational functions associated with each node; and $o \in N$ denotes the designated output node.

During execution, the graph receives a task input $x$, representing the instance of the current task. Nodes are executed in topological order. For a node $n$, let $\text{pre}(n)$ denote the set of its immediate predecessors. Node $n$ receives the concatenated outputs from its predecessors:
\begin{equation}
z_n = \{ f_v(z_v, x) : v \in \text{pre}(n) \}
\end{equation}
The node then executes its computational function $f_n(z_n, x)$ and passes the result to its successors. The final output of the graph is the result produced by the output node:
\begin{equation}
\hat{y} = G(x) = f_o(z_o, x)
\end{equation}

In a multi-agent scenario, each agent is itself a computational graph. Given a set of $K$ agents $\{ G_k = (N_k, E_k, F_k, o_k) \}_{k=1}^{K}$, a composite swarm graph is constructed by introducing cross-agent communication edges. Let:
\begin{equation}
N' = \bigcup_{k} N_k, \quad E' = \bigcup_{k} E_k, \quad F' = \bigcup_{k} F_k
\end{equation}
where $N'$, $E'$, and $F'$ represent the union of all nodes, internal edges, and computational functions, respectively. We introduce a set of cross-agent communication edges $E \subset \bigcup_{i \neq j} N_i \times N_j$, and define the complete edge set $E_E = E' \cup E$. The composite graph for the swarm is then:
\begin{equation}
G_E = (N', E_E, F', o')
\end{equation}
where $o'$ is the designated output node for the composite task.

\subsection{Topology Optimization Objective}

Within this framework, the problem of optimizing the LLM swarm structure reduces to selecting an optimal set of communication edges $E$. Given a task $\tau$ and a corresponding utility function $u_{\tau}(G_E)$, which typically depends on the correctness of the final task output, the objective of LLM swarm topology optimization is:
\begin{equation}
\max_{E} u_{\tau}(G_E)
\end{equation}
Let $d$ be the number of potential communication edges. The number of possible topologies is $2^d$, making direct search in the discrete space computationally prohibitive. To address this, we introduce a parameterized probability distribution $D_\theta$ to represent the sampling strategy for graph structures, where $\theta$ denotes the probability parameters for sampling each potential edge. The optimization is then formulated as maximizing the expected utility:
\begin{equation}
\arg \max_{\theta \in \Theta} \mathbb{E}_{G' \sim D_\theta} [u_{\tau}(G')]
\end{equation}
Under the REINFORCE framework, the gradient estimate is given by:
\begin{equation}
\nabla_\theta \mathbb{E}_{G'\sim D_{\theta}} [u_{\tau}(G')] \approx \frac{1}{M} \sum_{i=1}^{M} u_{\tau}(G_i) \nabla_{\theta} \log p_{\theta}(G_i), \quad G_i \sim D_{\theta}
\end{equation}
For each sampled graph $G_i \sim D_{\theta}$, its probability is computed sequentially ensuring acyclicity:
\begin{equation}
p_{\theta}(G_i) = \prod_{j=1}^{d} \begin{cases} 
\theta_j & \text{if } \big(N, E \cup (\{ e_k\}_{k=1}^{j-1} \cap \mathcal{E}) \cup \{e_j\}\big) \text{ is a DAG}, \\
0 & \text{otherwise.}
\end{cases}
\end{equation}

\subsection{External Context Score (ECS) Computation}

We adopt the External Context Score (ECS) from the ReDeEP framework~\cite{sun2024redeep}. 
To reduce computational cost while focusing on the layers most directly related to the final generation behavior, we compute ECS on a deterministic subset of late-layer attention heads. Specifically, we select the final 5\% of model layers and 10\% of attention heads within each selected layer. Let the model have $L$ layers, each containing $H$ attention heads. We define
\begin{equation}
M_E = \lceil 0.05L \rceil,
\end{equation}
and the selected layer set as
\begin{equation}
\mathcal{L}_{\text{ECS}} = \{ L-M_E+1, \dots, L \}.
\end{equation}
For each selected layer $l \in \mathcal{L}_{\text{ECS}}$, we deterministically select a subset of attention heads
\begin{equation}
\mathcal{H}_{\text{ECS}}^{l} \subseteq \{1, \dots, H\},
\end{equation}
with $|\mathcal{H}_{\text{ECS}}^{l}| = \lceil 0.1H \rceil$. The final ECS selection set is therefore
\begin{equation}
\mathcal{S}_{\text{ECS}} =
\{(l,h) \mid l \in \mathcal{L}_{\text{ECS}}, h \in \mathcal{H}_{\text{ECS}}^{l} \}.
\end{equation}
Segment-level ECS uses only the tokenizer indices belonging to the target content segment; role prompts and separators are excluded from the segment boundary.

For a generated token $t_n$ and a selected head $h$ in layer $l$, we define:
\begin{equation}
E_{n}^{l,h} = 
\frac{e \cdot x_n^L}{\| e \| \, \| x_n^L \|}, 
\quad 
e = \frac{1}{| I_{n}^{l, h} |} \sum_{j \in I_n^{l,h}} x_j^L ,
\end{equation}
where $x_n^L$ denotes the hidden state of token $t_n$ at the final layer $L$; $I_n^{l,h}$ denotes the set of context tokens that receive the highest attention weights from head $h$ in layer $l$ for token $t_n$; and $e$ is the average representation of these attended context tokens.

For a generated sequence $r$, the final ECS is averaged over all generated tokens and all selected layer-head pairs:
\begin{equation}
E_r =
\frac{1}{|\mathcal{S}_{\text{ECS}}|}
\sum_{(l,h)\in \mathcal{S}_{\text{ECS}}}
\frac{1}{|r|}
\sum_{t_n \in r}
E_n^{l,h}.
\end{equation}
ECS characterizes the model's reliance on external contextual information during generation by analyzing internal representations and attention flow, estimating the correlation between the generated output and the provided context.

\subsection{Parametric Knowledge Score (PKS) Computation}

We adopt the Parametric Knowledge Score (PKS) from ReDeEP~\cite{sun2024redeep}. Since knowledge editing literature indicates that factual knowledge is predominantly localized in the later feed-forward network (FFN) layers, we restrict PKS computation to the final 10\% of model layers. Let $M=\lceil 0.1L\rceil$. The deterministic layer selection set is defined as:
\begin{equation}
\mathcal{S}_{\text{PKS}} = \{ l \mid l \in \{L-M+1, \dots, L\} \}
\end{equation}

For a token $t_n$ and a layer $l \in \mathcal{S}_{\text{PKS}}$, the token-level PKS is defined using the Jensen-Shannon Divergence (JSD) between the output distributions projected from the hidden states before and after the FFN block:
\begin{equation}
P_n^l = \text{JSD}\left( q(x_n^{\text{mid}, l}) \,\|\, q(x_n^l) \right)
\end{equation}
where the projection function $q$ maps a hidden state to a probability distribution over the vocabulary:
\begin{equation}
q(x) = \text{softmax}(\text{LogitLens}(x))
\end{equation}
Here, $x_n^{\text{mid}, l}$ denotes the residual stream representation immediately before the FFN block of layer $l$; $x_n^l$ denotes the representation immediately after the FFN block; $\text{LogitLens}(\cdot)$ denotes the mapping of hidden states to vocabulary logits via the unembedding matrix; and $\text{JSD}(\cdot)$ is the Jensen-Shannon divergence.

For a generated sequence $r$, the sequence-level PKS for layer $l$ is the average over all tokens:
\begin{equation}
P_r^l = \frac{1}{|r|} \sum_{t \in r} P_t^l, \quad l \in \mathcal{S}_{\text{PKS}}
\end{equation}
PKS quantifies the extent to which the model's parametric knowledge influences the output; a high PKS indicates a significant change in the output distribution induced by the FFN block, reflecting reliance on memorized information rather than purely contextual cues.

\subsection{Shared Inference and Metric Settings}

Unless otherwise specified, all LLM calls use temperature 1.0, top-p 1.0, and a maximum generation length of 512 new tokens. We sample one generation per prompt and do not use greedy decoding. Models are evaluated without quantization. ECS and PKS are computed on the complete generated token sequence and averaged over all generated tokens.

\section{Detailed Experimental Setup for Behavioral Probe Analysis}
\label{app:ch3_details}

This appendix provides experimental details and reproducibility information
for the analysis of ECS and PKS as behavioral probes across node-level,
multi-source aggregation, chain-reasoning, and GAIA tool-augmented settings,
as described in Section~3 of the main paper.

\subsection{Node-Level Response Experiment}
\label{app:ch3_node_details}

\textbf{Dataset.} The node-level experiment employs all 1,513 cases from the \texttt{celebrity\_relations} dataset in the reversal\_curse repository~\cite{berglund2023reversal}. This dataset contains kinship questions about public figures (e.g., \textit{``Who is [Person A]'s mother?''}) where the correct answer is a named entity. For each question, four context conditions are constructed as described in Section~3.1 of the main paper: \textit{bubble}, \textit{rel}, \textit{oracle}, and \textit{answer}. All four conditions are evaluated for every question, and Figure~\ref{fig:node_level} aggregates scores at the generation level across prompt types. The exact phrasing of each context fragment follows the templates provided in the original reversal\_curse benchmark.

\textbf{Context Construction.} The four context conditions are designed to form an ordered usefulness gradient. The \textit{bubble} and \textit{rel} contexts are generated by an LLM. The \textit{bubble} context is a short, self-contained paragraph unrelated to the question, while the \textit{rel} context provides general background about the person mentioned in the question but excludes family-relation information. In contrast, \textit{oracle} and \textit{answer} contexts are constructed directly from the ground-truth relation: \textit{oracle} provides inverse relational evidence that supports reverse inference, and \textit{answer} directly states the target relation.

The prompt used to generate \textit{bubble} contexts is:
\begin{quote}
\small
\begin{verbatim}
You will be given a question. Your task is NOT to answer it.

Instead, generate a short paragraph (20-30 words) that is completely
unrelated to the question.

Requirements:
- Do NOT answer the question
- Do NOT mention or reference any part of the question
- Do NOT use similar keywords or topics from the question
- The paragraph should be self-contained and on a totally different topic

Question:
\end{verbatim}
\end{quote}

The prompt used to generate \textit{rel} contexts is:
\begin{quote}
\small
\begin{verbatim}
You will be given a question about a well-known person.

Your task is to generate a short paragraph (20-30 words) that provides
general background information about the person mentioned in the question.

Requirements:
- The paragraph should read like a neutral biographical description
- Focus on aspects such as career, achievements, public image, or general
  life background
- Do NOT include any information about family relationships (e.g., parents,
  children, siblings)
- Do NOT answer or help answer the question
- If you are not certain about factual details, you may generate plausible
  but realistic information

The goal is to produce text that appears informative and relevant to the
person, but is not useful for answering the question.

Question:
\end{verbatim}
\end{quote}

Table~\ref{tab:context_condition_examples} shows one representative example.
\begin{table}[t]
\centering
\small
\caption{Example context conditions for the node-level response experiment.}
\label{tab:context_condition_examples}
\begin{tabularx}{\linewidth}{>{\raggedright\arraybackslash}p{1.8cm} X}
\toprule
\textbf{Field} & \textbf{Text} \\
\midrule
Question & Who is Kate Mara's father? \\
\textit{bubble} & The rural landscape bloomed with vibrant wildflowers, each petal soaking up the sun's warmth. Farmers tended to their fields, hopeful for a bountiful harvest under the clear blue sky. \\
\textit{rel} & Kate Mara is an American actress renowned for roles in film and television, including ``House of Cards'' and ``Captain America: The First Avenger.'' She is also recognized for her fashion industry involvement and philanthropic work. \\
\textit{oracle} & Kate Mara is Timothy Christopher Mara's Child. \\
\textit{answer} & Timothy Christopher Mara is Kate Mara's father. \\
\bottomrule
\end{tabularx}
\end{table}

\textbf{Model Configuration.} Experiments were conducted on two models: \textit{Qwen-2.5-32B-Instruct} and \textit{Llama-3-70B-Instruct}. The figures presented in the main paper correspond to results from Qwen-2.5-32B-Instruct; identical qualitative conclusions were obtained with Llama-3-70B-Instruct. ECS and PKS are computed per token and averaged over the generated sequence, following the method of~\cite{sun2024redeep} and the shared settings above.

\subsubsection{Statistical Testing}
\label{app:ch3_node_stats}
For Figure~\ref{fig:node_level}, we evaluate monotonic association across the ordered relevance levels \textit{bubble}, \textit{rel}, \textit{oracle}, and \textit{answer} using a paired permutation Spearman test across the 1,513 questions. To preserve the paired structure, permutations are performed within each question by shuffling the four condition labels while keeping the four metric values associated with that question fixed. We also report adjacent-condition Cliff's $\delta$ with bootstrap confidence intervals; positive values indicate that the later condition in the ordered pair has larger metric values. Table~\ref{tab:fig1_statistical_summary} reports the statistics for the plotted prompt set.

\begin{table}[t]
\centering
\small
\caption{Statistical summary for the node-level response experiment in Figure~\ref{fig:node_level}. Mean adjacent Cliff's $\delta$ averages the three ordered comparisons \textit{bubble}$\rightarrow$\textit{rel}, \textit{rel}$\rightarrow$\textit{oracle}, and \textit{oracle}$\rightarrow$\textit{answer}. Confidence intervals are 95\% bootstrap intervals, and $p$-values are from paired permutation Spearman tests.}
\label{tab:fig1_statistical_summary}
\begin{tabular}{lcccc}
\toprule
\textbf{Metric} & \textbf{Mean adj. $\delta$} & \textbf{95\% CI} & \textbf{Spearman $\rho$} & \textbf{$p$} \\
\midrule
ECS & 0.624 & [0.610, 0.640] & 0.857 & $<2\times10^{-4}$ \\
PKS & -0.157 & [-0.170, -0.143] & -0.308 & $<2\times10^{-4}$ \\
\bottomrule
\end{tabular}
\end{table}

The pairwise adjacent comparisons further show that ECS increases reliably at every step of the relevance ordering. Cliff's $\delta$ is 0.772 for \textit{bubble}$\rightarrow$\textit{rel} (95\% CI [0.750, 0.794]), 0.972 for \textit{rel}$\rightarrow$\textit{oracle} ([0.965, 0.977]), and 0.129 for \textit{oracle}$\rightarrow$\textit{answer} ([0.088, 0.171]). PKS exhibits a less strictly local pattern but a significant overall decline with relevance: the largest decreases occur from \textit{rel}$\rightarrow$\textit{oracle} ($\delta=-0.211$, 95\% CI [-0.250, -0.171]) and \textit{oracle}$\rightarrow$\textit{answer} ($\delta=-0.294$, 95\% CI [-0.333, -0.255]).

\subsection{Multi-Source Aggregation Experiment}
\label{app:ch3_multi_source}

\textbf{Dataset and Task.} The expert-specialization experiment is conducted on the MMLU test set~\cite{hendrycks2020measuring} (Massive Multitask Language Understanding). We randomly sample 300 questions without balancing across topic groups and group subjects into seven broader topic categories to analyze how different role-defined experts are utilized across knowledge domains. The topic groups are summarized in Table~\ref{tab:mmlu_topic_groups}.

For each sampled question, multiple experts with predefined roles first generate independent responses, which are concatenated and provided to a downstream Critic node for evaluation. A Direct Answer Expert is always included as a domain-agnostic baseline that directly answers the question. The Critic subsequently passes its assessment to a Final Decision node that produces the final answer. We compute segment-level ECS for each expert-output segment in the concatenated Critic input, measuring how strongly the Critic behaviorally relies on different experts under each topic group.

\textbf{Expert Role Prompts.} Expert role prompts are taken from the MMLU prompt set in the open-source Agent Prune~\cite{zhang2024cut} implementation. Experts do not differ in external knowledge bases, retrieval access, or function-calling abilities; their domains are defined only through role prompts. Each expert receives a role description before the shared domain-assessment instruction. For example, the Doctor role is:
\begin{quote}
\small
\begin{verbatim}
You are a doctor and come up with creative treatments for
illnesses or diseases.
You are able to recommend conventional medicines, herbal remedies and
other natural alternatives.
You also consider the patient's age, lifestyle and medical history when
providing your recommendations.
\end{verbatim}
\end{quote}

The shared expert instruction is:
\begin{quote}
\small
\begin{verbatim}
Domain Assessment: First, determine if the question falls within your
core expertise as:

Condition A (Within Domain):
If the question is directly related to your field:
    Conduct a brief Step-by-Step reasoning.
    You MUST provide the final choice in the following JSON format:
    {"answer": "X"} (where X is A, B, C, or D).
    Don't forget to provide the reasoning step in your outputs

Condition B (Outside Domain):
If the question is NOT related to your field:
    DO NOT attempt to solve the problem or guess the answer.
    DO NOT use the JSON format {"answer": "X"}.
    Instead, provide a sophisticated, professional "fact or reflection" from
    your own domain that is entirely irrelevant to the question.
    Act as if you are preoccupied with your own research.
    For example, if you are a Biologist faced with a Math problem,
    discuss mitochondrial DNA or protein folding instead.
    DO NOT provide any JSON format answer in your outputs

Constraint: Your output must either contain the JSON answer (if in-domain)
or the irrelevant professional monologue (if out-of-domain).
Never mix the two.
\end{verbatim}
\end{quote}

\begin{table}[t]
\centering
\small
\caption{Topic grouping of MMLU subjects used in the expert-specialization analysis. The first column gives the aggregated topic group used on figure axes, and the second column lists the original MMLU subjects assigned to that group. This grouping enables cross-domain comparison of expert-conditioned ECS responses.}
\label{tab:mmlu_topic_groups}
\begin{tabularx}{\linewidth}{>{\raggedright\arraybackslash}p{3.4cm} X}
\toprule
\textbf{Topic Group} & \textbf{Included Subjects} \\
\midrule
Math & abstract algebra, college mathematics, elementary mathematics, high school mathematics, high school statistics, formal logic \\
Computer Science & college computer science, high school computer science, computer security, machine learning, electrical engineering \\
Physics & college physics, conceptual physics, high school physics, astronomy \\
Chemistry / Biology / Medicine & anatomy, clinical knowledge, college biology, college chemistry, college medicine, high school biology, high school chemistry, human aging, human sexuality, medical genetics, nutrition, professional medicine, virology \\
Law / Politics / History & high school US history, high school world history, high school European history, high school government and politics, international law, jurisprudence, professional law, security studies, US foreign policy, prehistory \\
Business / Economics & business ethics, econometrics, global facts, high school macroeconomics, high school microeconomics, management, marketing, professional accounting, public relations \\
Psychology / Philosophy / Social Science & high school psychology, professional psychology, philosophy, moral disputes, moral scenarios, logical fallacies, sociology, world religions \\
\bottomrule
\end{tabularx}
\end{table}

\textbf{Segment-Level ECS (Seg-ECS) Computation.} Within a single context window containing concatenated outputs from multiple upstream nodes, the input is partitioned into contiguous segments corresponding to each source. For each segment, ECS is computed independently following the standard token-level ECS method~\cite{sun2024redeep}, restricted to the token indices belonging to that segment. 

\textbf{Normalization.} To enable cross-question comparison, we apply z-score normalization \textit{within each question} following the same procedure used in the node pruning experiments of Section~5. Specifically, for a given question with segments $1,\dots,M$, let $\mu$ and $\sigma$ be the mean and standard deviation of the $M$ segment-level ECS values. The normalized score for segment $i$ is:
\begin{equation}
\tilde{\text{ECS}}_i = \frac{\text{ECS}_i - \mu}{\sigma}.
\end{equation}
This per-question standardization removes question-specific scale variations and highlights relative contribution differences among competing sources. The distributions shown in Figure~\ref{fig:expert_ecs} aggregate these normalized ECS values over all expert segments from the 300 sampled MMLU questions.
If the within-question standard deviation is zero, all normalized segment scores for that question are set to zero.

\subsubsection{Statistical Testing for Figure~\ref{fig:expert_ecs}}

To test whether domain-matched experts receive higher Critic reliance than non-matched experts, we use each question's Critic node as the paired unit. Of the 300 sampled MMLU questions, 268 have a clearly matched expert after excluding miscellaneous or hard-to-map categories. The Direct Answer Expert is excluded from this inferential comparison because it serves only as a visual baseline in Figure~\ref{fig:expert_ecs}; it is not treated as either a matched or non-matched domain expert.

For each valid paired unit, we compute
\begin{equation}
d = \overline{\text{ECS}}_{\text{matched expert}} -
\overline{\text{ECS}}_{\text{other non-Direct-Answer experts}},
\end{equation}
where the matched expert is the expert whose domain corresponds to the question category, and the other non-Direct-Answer experts are all remaining domain experts excluding Direct Answer. We then test the one-sided alternative $\overline{d}>0$ across paired units. Across $n=268$ valid pairs, the matched expert has a higher mean ECS than other non-Direct-Answer experts (0.3555 vs. 0.2820), with a mean paired difference of 0.0735 and a 95\% bootstrap confidence interval of [0.0683, 0.0786]. A one-sided Wilcoxon signed-rank test gives $p=1.65\times10^{-62}$, and a sign test shows that the matched expert is higher in 247 of 268 pairs ($p=1.97\times10^{-50}$).

\subsection{Chain Reasoning Experiment}
\label{app:ch3_chain_reasoning}

\textbf{Task Structure.} This appendix corresponds to the chain-reasoning analysis in Section~\ref{sec:chain_reasoning_analysis}. We manually select familiar high-level taxonomic groups so that the LLMs recognize the biological-classification context, then generate a controlled five-level hierarchy from NCBI Taxonomy records~\cite{schoch2020ncbi}. The resulting dataset contains 32 species instances with complete ranks for \textit{Phylum} (level 1), \textit{Class} (level 2), \textit{Order} (level 3), \textit{Family} (level 4), and \textit{Genus} (level 5). The goal at each step is to predict the correct genus given the accumulated taxonomic information up to the current level. In addition to the taxonomic context, the output produced by the previous chain position is passed to the next agent; thus each downstream position receives both the current taxonomic evidence and the preceding agent's generated response.

\textbf{Incremental and Truncated Conditions.}
\begin{itemize}[left=0pt, labelsep=5pt, topsep=0pt, partopsep=0pt, itemsep=2pt]
    \item \textbf{Full incremental condition:} At step $t \in \{1,\dots,5\}$, the model receives the taxonomic description of the $t$-th level (e.g., Step~1: \textit{``Phylum: Chordata''}; Step~2: adds \textit{``Class: Mammalia''}, etc.). For $t>1$, the prompt also includes the previous position's generated output. The model is prompted to predict the genus at every step.
    \item \textbf{Truncated condition with parameter $k$:} The model receives new taxonomic information only for steps $1,\dots,k$. For all subsequent steps $t > k$, the taxonomic context remains unchanged: the prompt provides the same level-$k$ taxonomic information \textit{without} advancing to deeper ranks. The previous position's generated output is still passed forward at every step, so the truncation removes only new taxonomic evidence, not inter-agent output propagation. The model is still required to make a genus prediction at each of these later steps, allowing observation of PKS behavior when no new taxonomic information is introduced.
\end{itemize}
PKS is computed as the sequence-averaged token-level PKS~\cite{sun2024redeep} for each generated prediction. Each species is evaluated under the full incremental condition and all truncated conditions ($k=1,2,3,4$). The trajectories shown in Figure~\ref{fig:chain_level} are averaged over all 32 species instances.

\subsection{GAIA Tool-Augmented Reasoning Pipeline}
\label{app:ch3_gaia_details}

The GAIA analysis in Section~\ref{sec:gaia_analysis} evaluates whether ECS and PKS remain informative in a realistic tool-augmented reasoning pipeline. Experiments use a manually filtered subset of GAIA Level-1 validation with video-modality instances removed. The resulting subset contains 52 instances. The system follows a Tree-of-Thought style route with four functional stages: \textit{GenerateQuery}, \textit{WebSearch}, \textit{FileAnalyse}, and \textit{CombineAnswer}. \textit{WebSearch} uses search API, \textit{FileAnalyse} extracts information from associated files when available, and \textit{CombineAnswer} synthesizes the collected evidence into the final answer.

The prompting setup follows the original GAIA design. The only modification is a minor instruction-following clarification in the \textit{WebSearch} stage: the original prompt states that when the ``information from web search is useless,'' the model should directly answer ``No useful information from WebSearch.'' We additionally specify that no extra analysis needs to be produced in this irrelevant-information case, so that the fixed phrase can be used more reliably as a relevance label.

For ECS analysis, we distinguish two segment-level analyses in the GAIA pipeline.
For Figure~\ref{fig:gaia_ecs}, the analyzed segments are the search-result segments returned by the function-calling tool during \textit{WebSearch}.
We label each search-result segment using the corresponding \textit{WebSearch} analysis output:
segments whose analysis output contains the fixed phrase ``No useful information from WebSearch''
are labeled irrelevant, while the remaining segments are treated as useful retrievals.
We then compute segment-level ECS during \textit{WebSearch} generation,
measuring the reliance of the generated \textit{WebSearch} analysis on each search-result segment.

For Figure~\ref{fig:gaia_seg_ecs}, we analyze evidence consumed by the downstream \textit{CombineAnswer} node.
Decision-criticality is defined branch-wise and estimated using two independent single-branch ablations.
In the \textit{WebSearch} ablation, we remove only \textit{WebSearch}-derived evidence while keeping \textit{FileAnalyse} evidence unchanged;
an instance is marked \textit{WebSearch}-critical only if the baseline answer is correct
and the answer becomes incorrect after removing \textit{WebSearch} evidence.
Similarly, in the \textit{FileAnalyse} ablation, we remove only \textit{FileAnalyse}-derived evidence while keeping \textit{WebSearch} evidence unchanged;
an instance is marked \textit{FileAnalyse}-critical only if the baseline answer is correct
and the answer becomes incorrect after removing \textit{FileAnalyse} evidence.
The branch that remains unchanged in each ablation is treated only as background context for that ablation,
and is not assigned criticality from the outcome of the other branch's removal.

In the original run, the baseline ToT pipeline answers 11 of 52 instances correctly,
corresponding to 21.2\% accuracy.
Removing \textit{WebSearch} evidence identifies three \textit{WebSearch}-critical cases
and reduces accuracy to 8 of 52 instances;
removing \textit{FileAnalyse} evidence identifies three \textit{FileAnalyse}-critical cases
and also reduces accuracy to 8 of 52 instances.
These two critical-evidence sets do not overlap, yielding six RAG-critical cases in total.

Under the threshold-based pruning analysis reported in the main text,
removing evidence above the approximately 40th-percentile ECS threshold
($\mathrm{ECS}\approx0.35$) reproduces the same 11/52 to 8/52 drop for each branch,
whereas removing evidence below the threshold leaves the number of correct answers unchanged.
For PKS analysis, we compare the \textit{CombineAnswer} stage before and after \textit{WebSearch} evidence ablation,
measuring whether the absence of retrieved evidence increases reliance on parametric knowledge.
The same prompt templates, tool-call formatting, and answer filtering rules are kept fixed between the original and ablated runs,
so that observed metric shifts can be attributed to the presence or absence of the ablated evidence.

\subsubsection{Statistical Testing for Figure~\ref{fig:gaia_tot}}
\label{app:ch3_gaia_stats}

For Figure~\ref{fig:gaia_tot}, unpaired ECS comparisons use label-permutation tests and percentile bootstrap confidence intervals for mean differences. Panel (a) uses the merged RAG-critical labeling described above, which contains six non-overlapping critical cases across \textit{WebSearch} and \textit{FileAnalyse}. The \textit{CombineAnswer} PKS comparison in panel (c) is paired across the 52 GAIA instances; we therefore report a paired bootstrap confidence interval, an exact two-sided Wilcoxon signed-rank test, and an exact two-sided sign test. Table~\ref{tab:fig3_statistical_summary} summarizes the results.

\begin{table}[t]
\centering
\scriptsize
\setlength{\tabcolsep}{3pt}
\caption{Statistical summary for the GAIA tool-augmented reasoning experiment in Figure~\ref{fig:gaia_tot}. Confidence intervals are 95\% bootstrap intervals for mean differences. Positive differences indicate larger ECS for critical RAG evidence in panel (a), larger ECS for relevant WebSearch segments in panel (b), and larger PKS after removing WebSearch evidence in panel (c).}
\label{tab:fig3_statistical_summary}
\resizebox{\linewidth}{!}{%
\begin{tabular}{lllll}
\toprule
\textbf{Panel} & \textbf{Comparison} & \textbf{Means} & \textbf{Mean diff.} & \textbf{Additional tests} \\
\midrule
Fig.~3a & RAG crit. vs. non-crit. ECS & 0.512 vs. 0.362 & $\Delta\mu=0.150$ [0.052, 0.246], $p=0.066$ & Cliff's $\delta=0.451$, $p=0.074$ \\
Fig.~3b & Rel. vs. irrel. ECS & 0.372 vs. 0.129 & $\Delta\mu=0.243$ [0.194, 0.285], $p<10^{-5}$ & Cliff's $\delta=0.974$, $p<10^{-5}$ \\
Fig.~3c & No-WS vs. WS PKS & 1.330 vs. 1.116 & $\Delta\mu=0.214$ [0.114, 0.309] & Wilcoxon $W=1118$, $p=4.75\times10^{-5}$; sign $k=41/52$, $p=3.59\times10^{-5}$ \\
\bottomrule
\end{tabular}%
}
\end{table}

\section{Conceptual Foundations and Interpretive Assumptions}
\label{app:ch4_interpretation}

This appendix clarifies the conceptual underpinnings of the heuristic interpretation presented in Section~4. We explicitly state the working assumptions that link ECS and PKS to information sources, and we provide a constructive counterexample illustrating why PKS decline alone does not guarantee correctness~\cite{zhao2025understanding,westphal2025generalized,zhang2026confidence}. \textbf{No formal mathematical proofs are claimed; the content below serves as an interpretive scaffold rather than a rigorous theory.}

\subsection{Heuristic Mapping from Behavioral Probes to Information Sources}

The core interpretive assumption of this work is a qualitative monotonic relationship
between the behavioral probes and the information sources they reflect:

\textbf{Assumption (Heuristic Monotonicity).}
For a given node $v$, the External Context Score (ECS) tends to increase
as the node depends more on the provided context $Z_v$.
Conversely, the Parametric Knowledge Score (PKS) tends to decrease
as the node depends less on its internal parametric knowledge $\Theta_v$.

This assumption is \emph{not} derived from first principles.
It is an empirical generalization motivated by the controlled experiments in Section~3:
\begin{itemize}[left=0pt, labelsep=5pt, topsep=0pt, partopsep=0pt, itemsep=2pt]
    \item ECS increases with input relevance (Figure~\ref{fig:node_level}a).
    \item PKS decreases as informative context accumulates (Figure~\ref{fig:chain_level}).
\end{itemize}

We do \emph{not} assert a precise functional form,
such as proportionality,
between these probes and any information-theoretic quantity.
The value of this assumption lies in providing a consistent heuristic interpretation
for examining information flow across interacting nodes
and for designing probe-guided variants of structure optimization.
It should be viewed as an empirical interpretation,
rather than a formal mechanism or theorem.

\subsection{Why PKS Decline Does Not Imply Correctness: A Constructive Counterexample}
\label{app:pks_to_acc}
A monotonic decrease in PKS reflects \emph{perceived} information sufficiency,
but it does not guarantee that the accumulated information is factually correct.
We illustrate this with a simple counterexample.

\noindent\textbf{Scenario.}
Consider a two-step reasoning chain where the final answer is a binary choice,
such as Yes/No.
The true answer is ``Yes.''
At Step~1, the model receives uninformative context
and produces a guess with moderate PKS.
At Step~2, the model receives a context that \emph{confidently but incorrectly} asserts ``No.''
Because the context is now specific and internally consistent,
the model's output becomes more determined by the misleading context.
As a result:
\begin{itemize}[left=0pt, labelsep=5pt, topsep=0pt, partopsep=0pt, itemsep=2pt]
    \item The model's parametric-knowledge dependence drops sharply
    $\Rightarrow$ \textbf{PKS declines}.
    \item The final output is ``No''
    $\Rightarrow$ \textbf{the answer is wrong}.
\end{itemize}

This scenario demonstrates that PKS decline reflects reduced parametric-knowledge dependence,
not the correctness of the context.
A declining PKS indicates that the model has become more settled under the provided context,
not that the answer is correct.
This is why, in our reasoning-path search experiment
(Section~\ref{sec:path_selection_case}),
we combine the PKS-based convergence probe with utility signals.

\subsection{GAIA ECS--PKS Behavioral Profiles as Empirical Support for Task-Dependent Regions}
\label{app:ch4_gaia}

This subsection provides additional context for the empirical observations on the GAIA dataset
referenced in Section~\ref{sec:metric_probe_interpretation}.
Using trajectories from the ToT reasoning pipeline,
we collect node-level ECS and PKS values for five functional log categories:
TotalQuery, WebQuery, FileAnalyse, WebAnalyse, and AdversarialAnswer.
Here, \textit{WebQuery} and \textit{WebAnalyse} are aliases for two internal substeps
of the same \textit{WebSearch} node:
\textit{WebQuery} denotes query construction before the function call,
while \textit{WebAnalyse} denotes summarization and analysis of the function-call results returned by \textit{WebSearch}.
They are not separate nodes in the swarm topology;
throughout the main paper and other appendices,
this stage is referred to uniformly as \textit{WebSearch}.

\begin{figure}[t]
  \centering
  \includegraphics[width=0.75\linewidth]{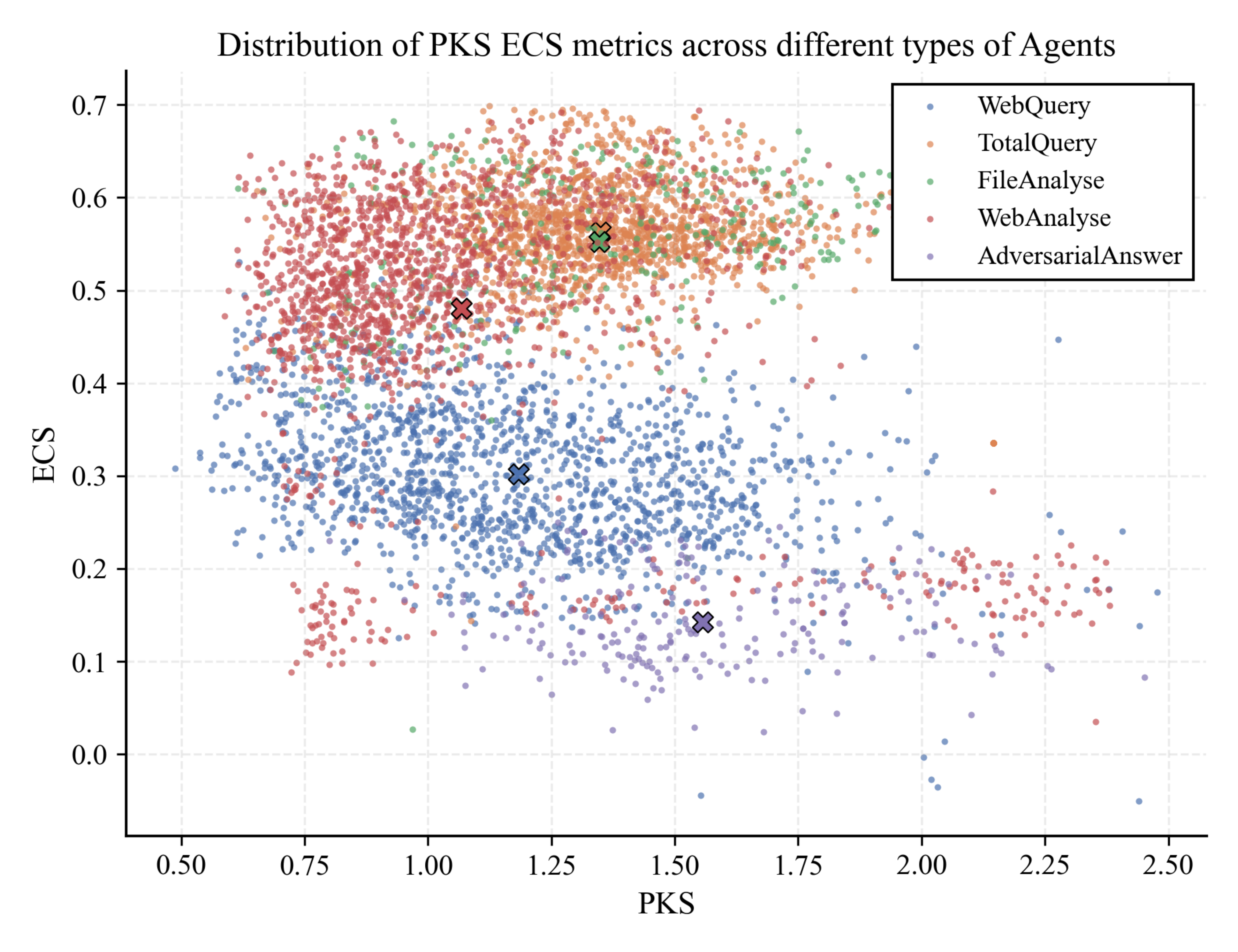}
  \caption{ECS--PKS behavioral profiles of functional categories in the GAIA dataset.
  The x-axis is PKS, the y-axis is ECS, colors denote functional log categories,
  and black crosses mark category centers.
  The profile clusters align with functional roles,
  supporting a task-dependent region $\Omega_{\text{task}}$ for probe-guided pruning.}
  \label{fig:gaia}
\end{figure}

As shown in Figure~\ref{fig:gaia},
functional categories associated with distinct roles occupy separable regions
in the ECS--PKS profile space:
\begin{itemize}[left=0pt, labelsep=5pt, topsep=0pt, partopsep=0pt, itemsep=2pt]
    \item \textbf{WebAnalyse} alias entries,
    corresponding to the analysis substep of the \textit{WebSearch} node,
    synthesize externally retrieved evidence and show high ECS with relatively low PKS.
    \item \textbf{AdversarialAnswer} nodes,
    which deliberately generate irrelevant responses,
    show low ECS and elevated PKS.
    \item \textbf{FileAnalyse} nodes,
    which process structured inputs and perform reasoning,
    show both high ECS and high PKS.
\end{itemize}

These clusters support the view that ECS--PKS behavioral profiles capture meaningful role-dependent patterns.
Moreover, outliers,
such as \textit{WebAnalyse} alias entries with unusually low ECS,
correspond to cases where the upstream retrieval is ineffective,
forcing the \textit{WebSearch} node to rely more on parametric knowledge.
This empirical pattern supports the use of a task-dependent region $\Omega_{\text{task}}$
and the corresponding probe-guided pruning criterion discussed in the main text.

\section{Detailed Setup for Probe-Guided Structure-Optimization Applications}
\label{app:ch5_details}

This appendix provides detailed task descriptions and task-specific reward definitions
for the two probe-guided structure-optimization applications in Section~5:
random-DAG node pruning and NCBI-derived taxonomic reasoning-path search.

\subsection{Random-DAG Node-Pruning Setup}
\label{app:node_pruning_setup}

The node-pruning experiment uses the MMLU development set filtered according to the GPTSwarm configuration, yielding 153 questions. All DA, LA, BA, and FD agents use the same backbone within a run, either Qwen-2.5-32B-Instruct or Llama-3-70B-Instruct. The graph contains two DA source nodes, two LA nodes, two BA nodes, and one FinalDecision (FD) sink node. DA nodes serve as fixed sources and FinalDecision serves as the fixed sink; beyond these input/output roles, we do not impose a chain, tree, or hand-designed intermediate topology. Instead, all admissible acyclic candidate connections among DA, LA, BA, and FD nodes are initialized with equal edge probability 0.5. This creates a controlled but nontrivial random-DAG search problem: the reward is the final 0/1 correctness signal, but many sampled paths pass through redundant LA/BA nodes or misleading intermediate combinations, while the target communication backbone is the direct DA$\rightarrow$FD route. Training uses the 200-iteration, learning-rate 0.1 protocol specified in the node-pruning reward definition below. At test time, edges are sampled according to the learned weights rather than deterministically thresholded.

\subsubsection{Node Roles and Output Protocol}

The four node labels used in the pruning graph correspond to \textit{DirectAnswer} (DA), \textit{LazyAnswer} (LA), \textit{BubbleAnswer} (BA), and \textit{FinalDecision} (FD). DA is the useful source type: it receives the question directly and is expected to solve it. LA is a redundant answer-preserving node: it receives an existing reference answer and is constrained to reproduce the same answer content with only minimal surface-level paraphrasing. BA is a distractor node: it produces confident but question-irrelevant content while still following the required answer format. FD is the sink node that aggregates the available upstream messages and produces the final multiple-choice decision.

The DA role prompt is intentionally minimal:
\begin{quote}
\small
\begin{verbatim}
You are a knowledgeable expert in question answering.
\end{verbatim}
\end{quote}

The LA instruction below summarizes the constraint used in implementation:
\begin{quote}
\small
\begin{verbatim}
You are a strict paraphrasing agent.

Your only task is to reproduce the reference answer exactly in meaning,
reasoning process, structure, and format, including the JSON answer section.
Do not add, remove, summarize, reorder, or reinterpret any content.
You may only make minimal surface-level paraphrases; if a paraphrase would
change the meaning or structure, copy the reference answer verbatim.
Output only the resulting answer.
\end{verbatim}
\end{quote}

The BA instruction is:
\begin{quote}
\small
\begin{verbatim}
Generate 1-2 confident sentences that are completely unrelated to the
question with the proper answer format.
\end{verbatim}
\end{quote}

All answer-producing nodes are required to provide reasoning together with a final answer, and the final answer must appear in JSON form:
\begin{quote}
\small
\begin{verbatim}
{"answer": "<FINAL_ANSWER>"}
\end{verbatim}
\end{quote}
Here, \texttt{<FINAL\_ANSWER>} must be one of the options in the question, usually a single letter. This shared output constraint ensures that DA, LA, BA, and FD remain format-compatible even when their informational roles differ, allowing topology optimization to distinguish useful direct-answer paths from redundant or irrelevant paths by their behavioral metric profiles rather than by parser failures.

\subsubsection{Random-DAG Sampling Procedure}
\label{app:dag_sample}
Candidate edges are defined by the input/output constraints above: the two DA nodes are source nodes, the FinalDecision node is the sink, and LA or BA nodes may be inserted through any directed connections that preserve acyclicity. All candidate edge logits are initialized so that the initial sampling probability is 0.5.

For each candidate edge $(u,v)$, we first check whether adding $u\rightarrow v$ preserves acyclicity. If so, the edge is sampled from a Bernoulli distribution with probability
\[
p_{uv}=\sigma(\theta_{uv}/\tau),
\]
where $\theta_{uv}$ is the learnable edge logit and $\tau$ is the sampling temperature. The realization log-probability accumulates $\log p_{uv}$ for selected edges and $\log(1-p_{uv})$ otherwise. After this first pass, we apply a predecessor-completion step to enforce the source-node constraint that the two DA nodes are the only sources. For every non-input node with zero in-degree, we randomly select one valid predecessor from its candidate incoming edges without violating acyclicity and force this edge to be included. The corresponding log-probability term is then replaced by the selected-edge term $\log p_{uv}$. This yields a realized DAG in which every non-input node has at least one predecessor, while only the two DA nodes can have zero in-degree.

\subsubsection{Computation of Edge Weight Metrics}
\label{app:ch5_weight_metric}

To quantify the learned communication topology, we measure the cumulative inbound edge weight to the final decision node (FD). Given a trained swarm graph with edge set $\mathcal{E}$, each directed edge $(u,v)\in\mathcal{E}$ is associated with a learned non-negative weight $w_{u,v}$ that reflects the communication strength between two agents.

For each intermediate node type $s \in \{\mathrm{DA}, \mathrm{LA}, \mathrm{BA}\}$, we aggregate all edges from nodes of type $s$ that terminate at FD:
\[
W_s=\sum_{(u \rightarrow \mathrm{FD}),\, u \in s} w_{u,\mathrm{FD}}.
\]

Here, $W_s$ represents the total contribution intensity of node type $s$ to the final decision process. Specifically, higher $W_{\mathrm{DA}}$ indicates stronger reliance on useful direct-answer agents, whereas lower $W_{\mathrm{LA}}$ and $W_{\mathrm{BA}}$ indicate that redundant or misleading communication paths have been successfully suppressed.

Following this definition, the total redundant communication weight is computed as:
\[
W_{\mathrm{redundant}} = W_{\mathrm{LA}} + W_{\mathrm{BA}}.
\]

All values reported in Table~\ref{tab:main_results} are averaged over evaluation samples after training convergence.

\subsection{NCBI-Derived Taxonomic Labyrinth for Path Selection}
\label{app:taxonomic_labyrinth}
\textbf{Task Structure.}
This appendix corresponds to the reasoning-path search experiment in Section~\ref{sec:path_selection_case}.
We use the same 32-instance NCBI-derived biological-classification set as the chain-reasoning analysis in Section~\ref{sec:chain_reasoning_analysis}.
Each instance contains a complete five-level taxonomy:
\textit{Phylum} (level 1), \textit{Class} (level 2), \textit{Order} (level 3), \textit{Family} (level 4), and \textit{Genus} (level 5).
Unlike the chain-reasoning analysis, where information is provided along a fixed sequence,
we convert each instance into a path-selection problem over five ordered reasoning positions.
The goal is to sample a reasoning path that follows the correct taxonomic progression and eventually predicts the correct genus.

\textbf{Path-Sampling Construction.}
For each instance, we create five ordered reasoning positions,
denoted as $\mathrm{pos}_1,\dots,\mathrm{pos}_5$,
together with a terminal node $\mathrm{T}$.
Each reasoning position corresponds to one taxonomic level and simulates one step of hierarchical refinement.
For example, $\mathrm{pos}_1$ provides the highest-level clue such as \textit{``Phylum: Chordata''};
$\mathrm{pos}_2$ should refine this with the class-level clue such as \textit{``Class: Mammalia''};
$\mathrm{pos}_3$ should add the order-level clue;
$\mathrm{pos}_4$ should add the family-level clue;
and $\mathrm{pos}_5$ corresponds to the genus-level clue.

At each iteration, a candidate reasoning path is sampled as a non-repeating Markov chain.
Sampling always starts from $\mathrm{pos}_1$.
Given the current position, the next node is sampled from the remaining unvisited reasoning positions and the terminal node
according to a softmax distribution over outgoing edge weights.
The sampling process stops once $\mathrm{T}$ is reached.
The terminal node does not participate in reasoning; it only serves as a stopping marker.
The final utility is computed from the answer produced by the last sampled reasoning position before $\mathrm{T}$.
Under this construction, the complete correct reasoning chain is
$\mathrm{pos}_1 \rightarrow \mathrm{pos}_2 \rightarrow \mathrm{pos}_3 \rightarrow \mathrm{pos}_4 \rightarrow \mathrm{pos}_5 \rightarrow \mathrm{T}$.

Taxonomic context is accumulated and passed downstream only when the sampled reasoning positions follow the correct order.
If the order is broken, downstream nodes receive only the initial highest-level taxonomic clue,
rather than the intermediate clues needed for stepwise refinement.
For example, a path that reaches $\mathrm{pos}_3$ without first visiting $\mathrm{pos}_2$
does not receive the accumulated class-level context;
similarly, a path that jumps from $\mathrm{pos}_1$ to $\mathrm{pos}_4$
cannot access the missing class- and order-level clues.
This construction makes the task harder when the sampled path deviates from the intended taxonomy,
while keeping the input format controlled across candidate paths.
Thus, only the fully ordered sampled chain reaches the complete taxonomic progression needed for reliable genus prediction.

\textbf{Optimization Setup.}
Optimization starts from edge weight $0.5$ and runs for $200$ iterations with learning rate $0.1$.
Each reported group uses three random seeds.
At each iteration, a non-repeating path is sampled using the softmax Markov-chain procedure described above.
The utility reward is $1$ if the final genus prediction from the last reasoning position before $\mathrm{T}$ is correct and $0$ otherwise.
For the probe-guided utility setting, optimization first uses utility-only rewards for $50$ warm-up iterations,
after which the PKS-based convergence probe is introduced.
The detailed PKS-based reward design is provided in Appendix~\ref{app:path_pks_reward}.

\subsection{Reward Function Definitions for Probe-Guided Structure Optimization}
\label{app:ch5_rewards}

This subsection provides the common reward-conversion procedure used by the two
probe-guided structure-optimization studies,
followed by the task-specific reward definitions for Section~5.1 and Section~5.2.
EMA smoothing reduces single-run probe noise,
robust normalization compares an edge- or node-level quantity against an appropriate comparison group,
and the piecewise function maps relative deviation to a bounded reward.
The semantic interpretation is determined by the input quantity $x_e$:
node pruning uses absolute ECS--PKS behavioral profiles,
while reasoning-path search uses the PKS drop as a convergence signal.

\subsubsection{Shared Reward Conversion}

\paragraph{Exponential Moving Average (EMA) Smoothing.}
For a generic metric $M \in \{\text{ECS}, \text{PKS}\}$ associated with edge $e$, the smoothed value at training iteration $t$ is computed as:
\begin{equation}
\tilde{M}_e^{(t)} = \beta \tilde{M}_e^{(t-1)} + (1-\beta) M_e^{(t)},
\end{equation}
where $\beta \in [0,1)$ is the smoothing coefficient. At $t=0$, $\tilde{M}_e^{(0)}$ is initialized with the first observed value $M_e^{(1)}$.

We linearly increase $\beta$ from 0.4 to 0.9 over the first 50 warmup iterations and keep it fixed at 0.9 afterwards.

\paragraph{Robust Normalization via Median and MAD.}
Inspired by the GRPO framework\cite{shao2024deepseekmath}, we standardize metrics within a comparison group using robust statistics. Let $\mathcal{G} = \{x_1, \dots, x_K\}$ denote a set of smoothed metric values. The normalized score for $x_i$ is:
\begin{equation}
z_i = \frac{x_i - \text{median}(\mathcal{G})}{c_s \cdot \text{MAD}(\mathcal{G})},
\end{equation}
where $\text{MAD}(\mathcal{G}) = \text{median}(\{|x_j - \text{median}(\mathcal{G})|\}_{j=1}^K)$ is the median absolute deviation, and $c_s = 1.4826$ ensures that $c_s \cdot \text{MAD}$ is a consistent estimator of the standard deviation for normally distributed data.

\paragraph{Piecewise Reward Function.}
Given a normalized score $z$, the edge-level reward is computed via a piecewise function that amplifies significant deviations from the group median while suppressing near-median variations:
\begin{equation}
g(z) = \begin{cases}
-r_{\min} - (r_{\max} - r_{\min}) \tanh\big(-(z + \tau)\big), & z < -\tau \\[4pt]
0, & |z| \le \tau \\[4pt]
r_{\min} + (r_{\max} - r_{\min}) \tanh\big(z - \tau\big), & z > \tau
\end{cases}
\label{eq:piecewise_reward}
\end{equation}
where $\tau > 0$ defines a ``dead zone'' around the median, and $r_{\min}, r_{\max} > 0$ control asymptotic penalty and reward magnitudes. The hyperbolic tangent $\tanh(\cdot)$ provides smooth saturation to extreme values.
We set $\tau=1.5$ and $r_{\min}=r_{\max}=0.5$ in all structure-optimization experiments.

\subsubsection{Section~5.1 Node-Pruning Reward}
\label{app:node_pruning_reward}

For a downstream node $n$ with predecessor set $\mathcal{P}(n)$, we first compute smoothed ECS values $\tilde{ECS}_{p_i \to n}$ for each incoming edge. Let $\mathcal{G}_n = \{\tilde{ECS}_{p_i \to n}\}$ be the local comparison group. The normalized ECS score $z^E_e$ is computed via robust normalization within $\mathcal{G}_n$.

We then apply a two-step reward design.

\paragraph{Step 1: ECS gating.}
If $z^E_e < 0$, the edge is treated as context-irrelevant for the downstream node. It receives only the ECS penalty and is excluded from the subsequent PKS normalization and reward computation. Only edges with $z^E_e \ge 0$ enter the second step.

\paragraph{Step 2: PKS gating.}
For edges that pass the ECS gate, we pool their smoothed absolute PKS values to form the global comparison group $\mathcal{G}_{\text{swarm}}$, and compute the normalized PKS score $z^P_e$. If $z^P_e < 0$, the edge is treated as a redundant aggregation path. It receives only the PKS penalty, and any positive ECS reward is ignored.

The resulting probe-derived pruning reward is:
\begin{equation}
R(e)=
\begin{cases}
\lambda_{\text{ECS}} g(z^E_e),
& z^E_e < 0, \\[1.5mm]
\lambda_{\text{PKS}} g(z^P_e),
& z^E_e \ge 0 \ \text{and}\ z^P_e < 0, \\[1.5mm]
\lambda_{\text{ECS}} g(z^E_e)
+
\lambda_{\text{PKS}} g(z^P_e),
& z^E_e \ge 0 \ \text{and}\ z^P_e \ge 0.
\end{cases}
\end{equation}
Here, $z^P_e$ is computed only for edges that pass Step 1. This design forces an edge to satisfy both criteria: context-irrelevant edges are filtered before PKS profiling, and low-PKS redundant paths cannot be retained solely by a positive ECS score.

In the final implementation, this probe-derived edge reward is combined with the output-driven task utility by direct addition. Let $U_{\tau}\in\{0,1\}$ denote the final-answer utility for task $\tau$, with $U_{\tau}=1$ when the sampled graph produces the correct output and $0$ otherwise. The reward used for updating the probability of edge $e$ is:
\begin{equation}
R_{\text{edge}}(e) = U_{\tau} + R(e).
\end{equation}
Thus, the ECS/PKS reward provides edge-local pruning guidance, while the 0/1 utility keeps the update tied to final output correctness. When a pruning loss is used, it is optimized in parallel with this reward-based edge-probability update and serves as a complementary structural regularizer rather than a replacement for either the output utility or the probe-derived reward.

Edge weights $w_e$ are updated via gradient ascent on cumulative reward. In our experiments, we use soft pruning only: no hard pruning threshold $\epsilon$ is applied, and evaluation samples edges stochastically according to the learned weights. Unless otherwise stated, $\lambda_{\text{ECS}}=\lambda_{\text{PKS}}=0.5$, training runs for 200 iterations, and the learning rate is 0.1.

\subsubsection{Section~5.2 Path-Selection Reward}
\label{app:path_pks_reward}

In the path optimization setting, the PKS reward is based on the PKS \textit{drop}
\begin{equation}
\Delta \tilde{P}_{u \to v} = \tilde{PKS}_u - \tilde{PKS}_v,
\end{equation}
where $u$ and $v$ are consecutive reasoning nodes. The two PKS values are
obtained before and after incorporating the information passed along $u\to v$,
then smoothed using the shared EMA procedure described above. Larger positive
drops indicate that the successor becomes less dependent on parametric memory
after receiving the upstream information. We normalize these drops locally among
successor edges of the same current node using the shared robust-normalization
rule, and convert the normalized score into an edge-level process reward:
\begin{equation}
R_{\text{PKS-drop}}(u\to v) =
g\!\left(z_{\Delta \tilde{P}_{u \to v}}\right).
\end{equation}

In the final \textit{Metric+Utility} implementation, this process reward is
combined with the output-driven utility by direct addition. Let
$U_{\tau}\in\{0,1\}$ denote the final-path utility for task $\tau$, with
$U_{\tau}=1$ if the sampled path produces the correct genus prediction and
$0$ otherwise. The edge reward used to update the edge probability is:
\begin{equation}
R_{\text{edge}}(u\to v)=
U_{\tau}+R_{\text{PKS-drop}}(u\to v).
\end{equation}
Thus, the PKS-drop term supplies edge-local process credit for convergence,
while the 0/1 utility anchors optimization to final output correctness.
In the \textit{Utility Only} baseline, the PKS-drop term is omitted and edge
updates rely only on $U_{\tau}$.
This defines the reward used in the Section~\ref{sec:path_selection_case} path-selection experiment.

\section{Details of Baseline-Inspired Loss Implementations}

This section provides the formal definitions of the two baseline-inspired losses used in our experiments. The GPTSwarm baseline follows the original configuration from~\cite{zhuge2024gptswarm} as a task-utility-based topology optimizer. For \textit{Agent Prune} and \textit{Agent Net}, which represent gradient-sensitivity pruning and trajectory-credit allocation, we do not claim strict reproduction of the full original systems; instead, we implement pruning- and trace-inspired losses under the same edge-level log-probability parameterization $\log p_e$ used by our swarm graph. This keeps the optimization interface comparable while preserving the core structural-credit intuition of each method.

\subsection{Agent-Prune- and Agent-Net-Inspired Losses}
\label{app:ch6_baseline}
\textbf{Agent Prune.}
The Agent-Prune-inspired loss constructs edge importance using gradient-based sensitivity of the task loss. Let $L_{\text{task}}$ denote the base policy loss. The gradient with respect to each edge log-probability is:
\begin{equation}
g_e = \frac{\partial L_{\text{task}}}{\partial \log p_e}.
\end{equation}

The edge importance is defined as:
\begin{equation}
I_e = \left| \log p_e \cdot g_e \right|.
\end{equation}

A sparsity-regularized reward is then constructed:
\begin{equation}
r_e^{\text{prune}} = \lambda_I I_e - \lambda_s,
\end{equation}
where $\lambda_I$ controls the importance scaling and $\lambda_s$ enforces uniform sparsity pressure across edges.
We set $\lambda_I=1.0$ and $\lambda_s=0.05$.

The policy is optimized via:
\begin{equation}
\mathcal{L}_{\text{prune}} = - \sum_{e} r_e^{\text{prune}} \log p_e.
\end{equation}

\vspace{0.5em}
\textbf{Agent Net.}
The Agent-Net-inspired loss assigns edge-level credit via a reward trace mechanism. Let $\mathcal{E}_t$ denote the set of activated edges at step $t$, and $R_t$ be the trajectory reward.

A decayed trace is maintained as:
\begin{equation}
T_e^{(t)} = \gamma T_e^{(t-1)} + \mathbb{I}(e \in \mathcal{E}_t)\cdot R_t,
\end{equation}
where $\gamma \in (0,1)$ is a decay factor.

The trace is normalized across edges:
\begin{equation}
\tilde{T}_e = \frac{T_e - \min(T)}{\max(T) - \min(T) + \epsilon}.
\end{equation}

The centered credit signal is defined as:
\begin{equation}
g_e = \tilde{T}_e - \frac{1}{2}.
\end{equation}

This signal modulates the policy update:
\begin{equation}
\mathcal{L}_{\text{net}} = - \sum_{e} \left( \text{stopgrad}(g_e) \cdot \log p_e \right),
\end{equation}
which is combined with the task objective:
\begin{equation}
\mathcal{L} = \mathcal{L}_{\text{task}} + \alpha \mathcal{L}_{\text{net}}.
\end{equation}
We set $\alpha=1.0$.

\section{Compute Resources}
The original experiments were run on a server with 8 NVIDIA A800 GPUs. Reproducing the full set of reported experiments is estimated to require approximately four days on comparable hardware. The main computational cost comes from repeated model inference, probe computation, and the probe-guided optimization runs. This estimate covers the original experiments and does not include preliminary exploratory runs or failed pilot experiments.

\section{Additional Evaluations and Robustness Analyses}
\label{app:additional_checks}
These supplementary evaluations retain the shared settings in Appendix~\ref{app:swarm_definition}, except for the explicitly varied decoding and probe selections.

\subsection{Simpler Signals and Configuration Robustness}

Table~\ref{tab:simpler_signals} compares simpler signals with ECS and
PKS on the same 6,052 single-node records (1,513 questions, four
conditions), with signal directions fixed before comparison.

\begin{center}
\begin{minipage}{\linewidth}
\centering
\captionof{table}{Comparison of simpler signals with ECS and PKS on
the same 6,052 single-node records.}
\label{tab:simpler_signals}
\small
\begin{tabular}{lcc}
\toprule
Signal & Ordering consistency & Aligned Spearman \\
\midrule
\multicolumn{3}{l}{\textit{ECS comparison}} \\
Raw attention mass & 0.4152 & -0.2897 \\
Length-normalized attention & 0.9755 & 0.9170 \\
ECS & 0.9147 & 0.8579 \\
\midrule
\multicolumn{3}{l}{\textit{PKS comparison}} \\
Entropy & 0.7437 & 0.4868 \\
Log-probability & 0.7368 & 0.4710 \\
PKS & 0.6499 & 0.2885 \\
\bottomrule
\end{tabular}
\end{minipage}
\end{center}

Length-normalized attention is competitive with ECS, whereas raw
attention mass performs substantially worse. Entropy and log-probability
discriminate these conditions more strongly than PKS. These single-node
results do not establish a universal ranking across swarm settings.

\begin{center}
\begin{minipage}{\linewidth}
\centering
\captionof{table}{Robustness to decoding settings. Stochastic results average seeds 42, 43, and 44 within each question--condition pair; questions remain the statistical units.}
\label{tab:decoding_robustness}
\small
\begin{tabular}{ccccc}
\toprule Decoding & ECS $\rho$ & ECS mean $\delta$ & PKS $\rho$ & PKS mean $\delta$ \\
\midrule
Greedy & 0.7995 & 0.5464 & -0.2487 & -0.1402 \\
$T=0.7$, three seeds & 0.8027 & 0.5537 & -0.2638 & -0.1457 \\
$T=1.0$, three seeds & 0.8603 & 0.6380 & -0.3162 & -0.1675 \\
\bottomrule
\end{tabular}
\end{minipage}
\end{center}
In Table~\ref{tab:decoding_robustness}, seeds 42/43/44 are averaged within each question--condition pair before testing; questions remain the statistical units ($p\le2\times10^{-4}$). Greedy and $T=0.7$ show a local oracle/answer ECS reversal, so only the overall association direction is robust. Six ECS subsets give $\rho\in[0.8369,0.8577]$; PKS over the last 4/7/13 FFN layers gives $\rho=-0.1614,-0.2955,-0.3852$, respectively (all $p<10^{-4}$). Absolute thresholds require recalibration.

\clearpage
\subsection{Paired Accuracy--Communication Evaluation}
Using the existing trained Qwen graphs without retraining or retuning, all methods are evaluated on the same 500 MMLU test questions. Probe-Joint obtains 80.2\% (401/500).
\begin{center}
\begin{minipage}{\linewidth}
\centering
\captionof{table}{Paired accuracy and communication comparison on 500 MMLU questions using the existing trained Qwen graphs. Probe-Joint achieves 80.2\% accuracy (401/500).}
\label{tab:paired_communication}
\small
\setlength{\tabcolsep}{3pt}
\renewcommand{\arraystretch}{1.2}
\begin{tabular}{>{\centering\arraybackslash}m{0.18\linewidth}>{\centering\arraybackslash}m{0.11\linewidth}>{\centering\arraybackslash}m{0.17\linewidth}>{\centering\arraybackslash}m{0.11\linewidth}>{\centering\arraybackslash}m{0.12\linewidth}>{\centering\arraybackslash}m{0.17\linewidth}}
\toprule
\shortstack{Baseline\\(accuracy)} & \shortstack{Accuracy\\difference\\(pp)} & \shortstack{95\% CI\\(pp)} & \shortstack{McNemar\\$p$} & \shortstack{Token\\reduction\\(\%)} & \shortstack{95\% CI\\(\%)} \\
\midrule
\shortstack{GPTSwarm\\81.2\%} & $-1.0$ & [$-3.2$, 1.0] & 0.458 & 29.7 & [27.7, 31.6] \\[3pt]
\shortstack{Agent Net\\81.2\%} & $-1.0$ & [$-3.2$, 1.0] & 0.458 & 41.4 & [39.9, 43.0] \\[3pt]
\shortstack{Agent Prune\\79.4\%} & $+0.8$ & [$-1.4$, 3.2] & 0.608 & 12.9 & [9.6, 16.0] \\[3pt]
\bottomrule
\end{tabular}
\par\smallskip
\footnotesize Accuracy differences are Probe-Joint minus baseline; pp denotes percentage points. Confidence intervals are 95\% intervals for paired differences and token reductions.
\end{minipage}
\end{center}
Table~\ref{tab:paired_communication} reports the paired estimates. Intervals including zero do not prove equivalence; token-reduction intervals are positive.

\subsection{Calibrated Workflow Evaluations}
For sparse expert selection, three questions per domain from the original MMLU question pool calibrate a fixed seg-ECS ranking before testing on the remaining questions in a seven-expert workflow.
\begin{center}
\begin{minipage}{\linewidth}
\centering
\captionof{table}{Sparse expert selection using a fixed, few-shot-calibrated seg-ECS ranking. Accuracy, expert calls, and tokens are reported per test question.}
\label{tab:sparse_selection}
\small
\begin{tabular}{cccc}
\toprule Policy & Acc. (\%) & Expert calls/question & Tokens/question \\
\midrule
All-7 & 77.46 & 7 & 5601.7 \\
ECS Top-5 & 76.94 & 5 & 4119.9 \\
Random-5 & 67.43 & 5 & 3929.0 \\
\bottomrule
\end{tabular}
\end{minipage}
\end{center}
Table~\ref{tab:sparse_selection} reports sparse selection. For conditional escalation, an independent 115-question pool supplies 100 calibration resamples, each with 10 base-workflow errors and 10 correct cases. A fitted PKS threshold gates an external reasoning Adapter on the same 300 held-out questions; Random Gate matches each threshold's call count.
\begin{center}
\begin{minipage}{\linewidth}
\centering
\captionof{table}{Conditional escalation to an external reasoning Adapter on 300 held-out questions. Incremental tokens are measured relative to the No-Adapter workflow.}
\label{tab:conditional_gate}
\small
\begin{tabular}{cccc}
\toprule Policy & Acc. (\%) & Mean Adapter calls/300 & Incremental tokens/question \\
\midrule
Random Gate & 79.95 & 102.09 & 613.74 \\
PKS Gate & 82.08 & 102.09 & 729.97 \\
Always Adapter & 85.67 & 300 & 1803.46 \\
\bottomrule
\end{tabular}
\end{minipage}
\end{center}
In Table~\ref{tab:conditional_gate}, incremental tokens include additional Adapter and FinalDecision input/output relative to No-Adapter. PKS gains 2.13 pp over Random Gate (question-level paired bootstrap 95\% CI [0.60,3.78], $p=0.0026$), at matched calls, not matched tokens. These workflows provide preliminary transfer evidence, not general optimization guarantees.

\clearpage
\subsection{Probe Cost}
For generation length $T$, context length $C$, hidden size $d$, vocabulary size $V$, selected ECS pairs $S_E$, selected PKS layers $S_P$, and $k$ retained context tokens/head, incremental costs are approximately $O(TS_E(C+kd))$ for ECS and $O(TS_PdV)$ plus $O(TS_PV)$ for PKS.
\begin{center}
\begin{minipage}{\linewidth}
\centering
\captionof{table}{Incremental probe-collection cost on six RTX 4090 GPUs. Overhead is relative to generation without probes; timing is in milliseconds per generated token.}
\label{tab:probe_latency}
\small
\begin{tabular}{ccc}
\toprule Mode & ms/token & Overhead \\
\midrule
Base & 84.546 & -- \\
ECS & 97.108 & 14.9\% \\
PKS & 110.247 & 30.4\% \\
ECS+PKS & 112.312 & 32.8\% \\
\bottomrule
\end{tabular}
\end{minipage}
\end{center}
Table~\ref{tab:probe_latency} uses Qwen2.5-32B, FP16, eager attention, six RTX 4090 GPUs, 16 ECS layer--head pairs in the final four layers and seven final PKS layers; medians follow two warm-ups and five measured runs. Joint collection shares one generation. Online PKS gating retains this overhead; comparison with fused-attention production inference may differ.
\newpage
\input{checklist.tex}

\end{document}

%% file: checklist.tex
\section*{NeurIPS Paper Checklist}

\begin{enumerate}

\item {\bf Claims}
    \item[] Question: Do the main claims made in the abstract and introduction accurately reflect the paper's contributions and scope?
    \item[] Answer: \answerYes{} 
    \item[] Justification: The abstract and introduction frame the paper as an empirical study of information flow in LLM swarms using process-level behavioral probes. The claims are limited to ECS/PKS as one concrete probe instantiation, their observed patterns in controlled swarm topologies and a realistic tool-augmented GAIA pipeline, and two controlled applications where these probes complement utility-driven structure optimization. We explicitly present the probe interpretation as heuristic, not as a formal decomposition of generation or a standalone optimization guarantee.

\item {\bf Limitations}
    \item[] Question: Does the paper discuss the limitations of the work performed by the authors?
    \item[] Answer: \answerYes{} 
    \item[] Justification: The paper includes a separate Limitations section. It states that the study is empirical and that ECS/PKS are used as heuristic behavioral probes rather than formal mechanisms. It also discusses the limited coverage of datasets and model families, the controlled nature of the optimization studies, and the additional computational cost of probe computation.

\item {\bf Theory assumptions and proofs}
    \item[] Question: For each theoretical result, does the paper provide the full set of assumptions and a complete (and correct) proof?
    \item[] Answer: \answerNA{}
    \item[] Justification: The paper does not present theoretical results such as theorems, lemmas, or formal proofs. The mathematical expressions in the paper define metrics, probe computations, and reward functions used in the empirical study.

\item {\bf Experimental result reproducibility}
    \item[] Question: Does the paper fully disclose all the information needed to reproduce the main experimental results of the paper to the extent that it affects the main claims and/or conclusions of the paper (regardless of whether the code and data are provided or not)?
    \item[] Answer: \answerYes{}
    \item[] Justification: The paper provides the information needed to reproduce the main empirical results at the level relevant to its claims. It describes the controlled topology experiments, the tool-augmented GAIA pipeline, the chain/path-selection task, model choices, probe computation procedures, ablation settings, reward definitions, and evaluation metrics. The appendix further specifies dataset construction, implementation details, and additional experimental definitions. Since the contribution is an empirical analysis and probe-guided application study rather than a new model or architecture, reproducibility is supported through detailed experimental and methodological descriptions.

\item {\bf Open access to data and code}
    \item[] Question: Does the paper provide open access to the data and code, with sufficient instructions to faithfully reproduce the main experimental results, as described in supplemental material?
    \item[] Answer: \answerYes{}
    \item[] Justification: The code and processed artifacts will be released at \url{https://github.com/HaRry-qaq/Swarmprobe}, together with instructions for reproducing the main experimental results. The paper and appendix describe the experimental settings, data construction, probe computations, rewards, and evaluation protocols. The release will include environment specifications, dependency versions, and commands for data preparation and experiment execution. It will also provide instructions for computing ECS and PKS and reproducing the main tables and figures.

\item {\bf Experimental setting/details}
    \item[] Question: Does the paper specify all the training and test details (e.g., data splits, hyperparameters, how they were chosen, type of optimizer) necessary to understand the results?
    \item[] Answer: \answerYes{}
    \item[] Justification: The paper specifies the experimental details necessary to understand the results. Since the work does not train a new model, the relevant details are the task construction, dataset filtering, model choices, topology and pipeline designs, probe computation procedures, ablation settings, reward definitions, and evaluation metrics. Additional implementation details and extended experimental definitions are provided in the appendix.

\item {\bf Experiment statistical significance}
    \item[] Question: Does the paper report error bars suitably and correctly defined or other appropriate information about the statistical significance of the experiments?
    \item[] Answer: \answerYes{}
    \item[] Justification: The paper provides statistical evidence for the quantitative results that support the main empirical claims, including confidence intervals and effect-size measures where applicable. Some analyses are qualitative; for these, the paper reports them as descriptive or exploratory rather than as standalone significance claims. The statistical procedures and additional quantitative summaries are described in the appendix.

\item {\bf Experiments compute resources}
    \item[] Question: For each experiment, does the paper provide sufficient information on the computer resources (type of compute workers, memory, time of execution) needed to reproduce the experiments?
    \item[] Answer: \answerYes{}
    \item[] Justification: The paper reports the compute resources used for the experiments in the appendix. The original experiments were run on a server with 8 NVIDIA A800 GPUs, and reproducing the full set of experiments is estimated to require approximately four days on comparable hardware. The appendix additionally reports probe-latency measurements on six RTX 4090 GPUs.
    
\item {\bf Code of ethics}
    \item[] Question: Does the research conducted in the paper conform, in every respect, with the NeurIPS Code of Ethics \url{https://neurips.cc/public/EthicsGuidelines}?
    \item[] Answer: \answerYes{}
    \item[] Justification: The research conforms to the NeurIPS Code of Ethics. The experiments use public or constructed datasets and evaluate LLM swarm behavior through model outputs, probe computations, and controlled analyses. The work does not involve human-subject experiments, private personal data, or deployment in high-risk decision-making settings. The camera-ready version uses the final conference format.

\item {\bf Broader impacts}
    \item[] Question: Does the paper discuss both potential positive societal impacts and negative societal impacts of the work performed?
    \item[] Answer: \answerYes{}
    \item[] Justification: The paper discusses positive impacts through improved diagnosis, interpretability, and efficiency of LLM swarm and tool-augmented reasoning systems, and also notes potential negative impacts from misuse of more capable automated agent systems and over-interpretation of heuristic probes as definitive explanations or safety guarantees.

\item {\bf Safeguards}
    \item[] Question: Does the paper describe safeguards that have been put in place for responsible release of data or models that have a high risk for misuse (e.g., pre-trained language models, image generators, or scraped datasets)?
    \item[] Answer: \answerNA{}
    \item[] Justification: The paper does not release a new pretrained model, image generator, or high-risk scraped dataset. The planned release is limited to code, configurations, controlled task data, and processed artifacts for reproducing the empirical analyses. These materials do not provide new generative capabilities or unsafe scraped content, so safeguards for high-risk model or dataset release are not applicable.

\item {\bf Licenses for existing assets}
    \item[] Question: Are the creators or original owners of assets (e.g., code, data, models), used in the paper, properly credited and are the license and terms of use explicitly mentioned and properly respected?
    \item[] Answer: \answerYes{}
    \item[] Justification: The paper cites the original sources of the datasets, models, and methods used in the experiments, including GAIA, MMLU, NCBI Taxonomy, the LLM backbones, and the ReDeEP probe definitions. Asset-specific licenses and terms of use will be documented in the appendix and release package. Any released materials are limited to code, configurations, controlled task data, and processed artifacts needed for reproducibility, with attribution and license information included in the release package.

\item {\bf New assets}
    \item[] Question: Are new assets introduced in the paper well documented and is the documentation provided alongside the assets?
    \item[] Answer: \answerYes{}
    \item[] Justification: The planned release contains code, configurations, and processed artifacts. The experiments use existing benchmarks, public model backbones, and constructed or processed experimental data described in the paper and appendix. These materials will be released at \url{https://github.com/HaRry-qaq/Swarmprobe} with accompanying documentation and license information.

\item {\bf Crowdsourcing and research with human subjects}
    \item[] Question: For crowdsourcing experiments and research with human subjects, does the paper include the full text of instructions given to participants and screenshots, if applicable, as well as details about compensation (if any)? 
    \item[] Answer: \answerNA{}
    \item[] Justification: The paper does not involve crowdsourcing experiments or research with human subjects. The experiments are conducted using existing benchmarks, constructed or processed task data, and LLM-generated outputs, without collecting data from human participants or crowd workers.

\item {\bf Institutional review board (IRB) approvals or equivalent for research with human subjects}
    \item[] Question: Does the paper describe potential risks incurred by study participants, whether such risks were disclosed to the subjects, and whether Institutional Review Board (IRB) approvals (or an equivalent approval/review based on the requirements of your country or institution) were obtained?
    \item[] Answer: \answerNA{}
    \item[] Justification: The paper does not involve crowdsourcing or research with human subjects. No data are collected from study participants, and the experiments are conducted using existing benchmarks, constructed or processed task data, and LLM-generated outputs. Therefore, IRB approval or equivalent human-subject review is not applicable.

\item {\bf Declaration of LLM usage}
    \item[] Question: Does the paper describe the usage of LLMs if it is an important, original, or non-standard component of the core methods in this research? Note that if the LLM is used only for writing, editing, or formatting purposes and does \emph{not} impact the core methodology, scientific rigor, or originality of the research, declaration is not required.
    \item[] Answer: \answerYes{}
    \item[] Justification: The paper describes the use of LLMs as core components of the studied swarm systems. LLMs are used as reasoning nodes in controlled topologies, tool-augmented GAIA pipelines, and probe-guided optimization studies. The paper specifies the model backbones used in the experiments and describes how LLM outputs, internal computations, and probe scores are analyzed. LLMs used only for writing or editing assistance, if any, are not part of the scientific methodology.

\end{enumerate}